\documentclass[aps,reprint,groupedaddress,superscriptaddress,pra]{revtex4-2}

\usepackage[utf8]{inputenc}
\usepackage{amsmath}
\usepackage{amsfonts}
\usepackage{amssymb}
\usepackage{bm}
\usepackage{color}
\usepackage{tcolorbox}
\usepackage{graphicx}
\usepackage{soul}
\usepackage{physics}
\usepackage{mathrsfs}
\usepackage{multirow}
\usepackage{xcolor}
\usepackage{mathtools}
\usepackage{lipsum}
\usepackage{comment}
\usepackage{booktabs}
\usepackage{textcomp}

\usepackage[colorlinks,citecolor=blue,linkcolor=red,anchorcolor=blue,urlcolor=blue]{hyperref}

\begin{document}
\title{Finite-Time Thermodynamics of an Autonomous Information Machine}
        
\author{Wanyan Chen}
\altaffiliation{These authors contributed equally to this work.}
\affiliation{School of Physics and Astronomy, Beijing Normal University, Beijing, 100875, China}
\affiliation{Key Laboratory of Multiscale Spin Physics (Ministry of Education), Beijing Normal University,Beijing 100875, China}

\author{Miao Chen}
\altaffiliation{These authors contributed equally to this work.}
\affiliation{School of Physics and Astronomy, Beijing Normal University, Beijing, 100875, China}
\affiliation{Key Laboratory of Multiscale Spin Physics (Ministry of Education), Beijing Normal University,Beijing 100875, China}

\author{Yu-Han Ma}
\email{yhma@bnu.edu.cn}
\affiliation{School of Physics and Astronomy, Beijing Normal University, Beijing, 100875, China}
\affiliation{Key Laboratory of Multiscale Spin Physics (Ministry of Education), Beijing Normal University,Beijing 100875, China}

\begin{abstract} 
While externally driven information engines are well understood, the thermodynamic constraints of their autonomous counterparts remain an open question. Here, we investigate the finite-time operation of an autonomous machine functioning as both an information eraser and a refrigerator, revealing that its irreversibility is bounded by the transient information geometry. Beyond steady-state boundaries, we map the landscape of optimal operation times across both functional modes, uncovering a unique synergistic regime where erasure power $P$ and efficiency $\eta$ increase simultaneously. Fundamentally, this performance is governed by a trade-off relation, $(1-\eta)P/\eta \le Dv$, where $v$ is the operational speed and $D$ denotes an information-geometric distance. Our findings pave the way for optimizing fast autonomous information-energy conversion.
\end{abstract}

\maketitle
\textit{Introduction}.---The integration of information theory and thermodynamics has profoundly reshaped our understanding of nonequilibrium processes~\citep{parrondo_thermodynamics_2015, Gour2015,Sagawa2012,Horowitz2014,Wolpert2019}. At the heart of this connection lies Landauer's principle, which establishes the fundamental thermodynamic cost of information processing~\citep{Landauer1961,Bennett1982,Brut2012}. Recently, the field has shifted its focus from quasistatic bounds to the realm of finite-time thermodynamics~\citep{Bjarne2022,kosloff2014quantum,qiu20roadmap}. For simple information erasure driven by explicit external protocols, the finite-time Landauer principle reveals a clear trade-off where operating at higher speeds requires dissipating more heat~\citep{Proesmans2020,Zhen2021, Lee2022, Dago2022,2022Ma,2022Van}. %In such externally driven processes, this implies that the rate of information erasure (power) can theoretically diverge to infinity, provided one supplies an unbounded energetic cost.

Beyond simple erasure processes, studies on diverse information-thermodynamic cycles~\citep{Quan2006, Toyabe2010,Cai2012,Mandal2012,2013PRLMQJ,Browne2014,Zulkowski2014,Tushar2021,Danageozian2022,Fadler2023,2024Zhou,Erdman2025,aguilar2025power} have advanced rapidly, uncovering critical performance bounds such as efficiency at maximum power and trade-offs among power, efficiency, and stability. For instance, our prior work~\citep{2024Zhou} on finite-time quantum Szilard engines demonstrated that information-recording dynamics strictly bounds energy conversion. However, relying on idealized, nonautonomous frameworks leaves a critical gap. By assuming explicit externally driven cyclic protocols and treating operational time as a freely tunable parameter, existing models decouple the system from the backaction and the inevitable thermodynamic cost of the controller. This idealization masks the true physical limits of practical operation, implying that the operational power could unphysically diverge to infinity given unbounded energy.

In contrast, natural and synthetic molecular machines operate autonomously under steady thermodynamic forces~\citep{Kosloff2013,Barato2014,McGrath2017,Amano2022}. In these self-contained systems, explicit time-dependent control is entirely replaced by internal stochastic transitions. Consequently, the operational time emerges as a dynamically restricted variable rather than a free parameter, imposing an inevitable thermodynamic cost of autonomy~\citep{2025Jarzynskia}. Notably, Mandal, Quan, and Jarzynski provided an exactly solvable model of an autonomous Maxwellian machine (the MQJ model)~\cite{2013PRLMQJ}. This seminal work, along with subsequent studies~\citep{cao2015,Cao2023Chem, Bao2023}, has successfully addressed the machine's steady-state performance boundaries and transient startup process. However, the finite time thermodynamic bounds governing its continuous operation remain uncharted. This leaves a central open question: Devoid of external driving, do intrinsic dynamics impose strict physical limits on the efficiency and power of autonomous information-energy conversion, or can the performance still theoretically diverge during infinitesimally fast interactions?

To directly answer this inquiry, we systematically investigate the finite-time thermodynamics of the MQJ model~\cite{2013PRLMQJ} operating in its functional steady state. By mapping the finite-time operational phase space of this autonomous system and deriving information-geometric bounds on the machine's entropy production, we demonstrate that intrinsic dynamics impose fundamental thermodynamic speed and performance limits on information-energy conversion. This provides necessary insights for designing high-frequency synthetic machines and understanding biological networks under strict temporal constraints.

\begin{figure*}
    \centering
    \includegraphics[width=1\linewidth]{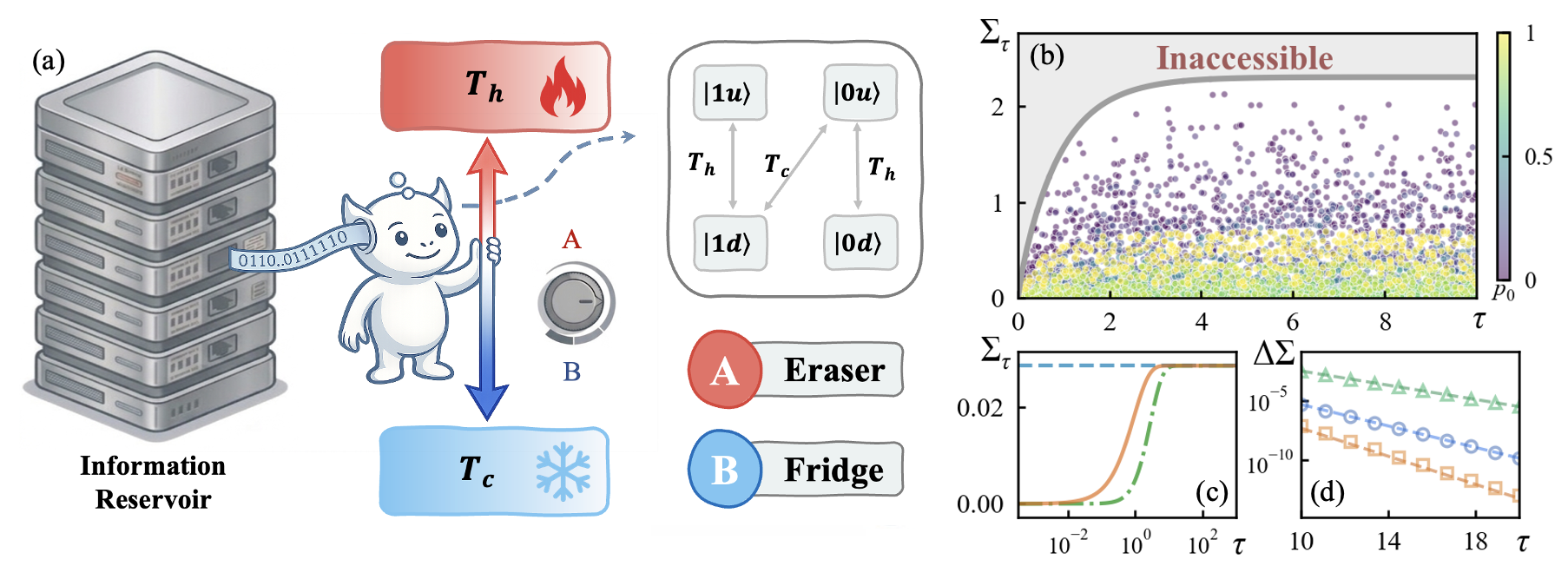}
    \caption{(a) Schematic of the autonomous information machine. A ``demon" processes incoming bits to direct energy transfer between two heat reservoirs, allowing the system to operate as either a refrigerator or an information eraser. (b) Total entropy production $\Sigma_\tau$ as a function of interaction time $\tau$. The color gradient indicates the initial probability $p_0$ of the incoming bits over the parameter range $p_0 \in [0,1]$, $\omega \in [0.2, 0.8]$, $\sigma < \omega$. (c) Bounds on finite-time entropy production. For a fixed parameter set $\Lambda=\{0.3,0.5,1,0.5\}$, $\Sigma_{irr}(\tau)$ (solid line) is tightly constrained between the instantaneous $D_{KL}(p_0 || p_\tau)$(dash-dotted line) and its asymptotic limit $D_{KL}(p_0 || p_\infty)$(dashed line). (d) Semi-logarithmic plot of $\Delta \Sigma$ versus $\tau$. The exponential relaxation is governed by $\lambda_3$, the dominant non-zero eigenvalue of the transition rate matrix. Markers denote numerical results for $\Lambda=\{0.1, 0.9, 0.5, 0.2\}$ (triangles), $\Lambda=\{0.2, 0.5, 1, 0.95\}$ (circles), and $\Lambda=\{0.2, 0.5, 2, 0.5\}$ (squares), while dashed lines indicate the corresponding theoretical decay rates. $\gamma=1$ throughout this paper.}
    \label{fig:Thermodynamic bound}
\end{figure*}

\textit{Tight dissipation bounds for an autonomous information machine.}---As illustrated in Fig. \ref{fig:Thermodynamic bound}(a), the MQJ model comprises a two-level demon ($D$) interacting with a stream of bits ($B$), and two thermal reservoirs at temperatures $T_h$ and $T_c$ ($T_c < T_h$). The demon is a two-state system with states $u$ and $d$, characterized by an energy gap $\Delta E = E_u - E_d > 0$. Its intrinsic transitions are driven by the hot reservoir, with transition rates $R_{u \to d}$ and $R_{d \to u}$ satisfying the detailed balance condition $R_{u \to d}/R_{d \to u} = e^{-\beta_h \Delta E}$, where $\beta_h = 1/(k_B T_h)$ and $k_B$ is Boltzmann's constant. To parametrize, we define
\begin{equation}
    R_{u \to d(d \to u)} = \gamma(1\pm\sigma), \quad \sigma = \tanh \frac{\beta_h \Delta E}{2} \label{eq:D-rate}
\end{equation}
where $\gamma > 0$ sets the characteristic transition rate of the demon. Besides these intrinsic transitions, the demon interacts sequentially only with the nearest bit on the tape over a finite time interval $\tau$, assuming the tape moves at a uniform speed and the bits possess no intrinsic dynamics of their own. Bit flips are strictly restricted to the cooperative joint transitions $0u \leftrightarrow 1d$, occurring solely via thermal contact with the cold reservoir. These rates are given by
\begin{equation}
    \quad R_{1d \to 0u(0u \to 1d)} = 1\pm\omega, \quad \omega = \tanh\frac{\beta_c \Delta E}{2} \label{eq:co-rate}
\end{equation}
where $\beta_c = 1/(k_B T_c)$. These rates are assembled to construct the continuous-time Markov transition rate matrix $\mathbf{R}$. To describe the coupled system statistically, the joint probability distribution is defined as $\mathbf{P} = (p_{0u}, p_{0d}, p_{1u}, p_{1d})^T$. The evolution of this system within the interaction interval $\tau$ is governed by the master equation $\dot{\mathbf{P}} = \mathbf{R} \mathbf{P}$. After a large number of interactions, the system relaxes into a periodic steady state. Here, the marginal distribution of the demon remains identical at the beginning and end of each interaction interval, while the outgoing bit stream settles into a well-defined, parameter-dependent configuration. Our subsequent analysis focuses on the finite-time thermodynamic constraints and performance within this periodic steady state.

During the interaction process, the demon acts as an energy conduit between the two reservoirs. If the fraction of $1$'s in the outgoing stream is greater than in the incoming stream, energy must be systematically transported from the cold reservoir to the hot reservoir. The average transformed energy per interval is
\begin{equation}
    \Delta Q = \Delta E (p_1^\prime - p_1) \label{dQ}
\end{equation}
where $p_1$ is the probability of $1$'s among the incoming stream, and $p_1^\prime$ is that of the outgoing stream. Accompanying this energy transfer, the information content of the bit stream undergoes a corresponding change. We use the Shannon entropy to quantify this change in information
\begin{equation}
    \Delta S_B = S_B(p_1^\prime)- S_B(p_1) \label{dSB}
\end{equation}
where $S_B(p) = - \sum_{i \in \{0, 1\}} p_i \ln p_i$. The total irreversible entropy production per interaction interval is therefore
\begin{equation}
    \Sigma_\tau = \Delta Q(\beta_h - \beta_c) + \Delta S_B
    \label{Entropyproduction}
\end{equation}
In the original formulation of this model~\cite{2013PRLMQJ}, it was established that the second law of thermodynamics mandates a simple non-negativity constraint, $\Sigma_\tau \ge 0$. However, our finite-time analysis in the following reveals that the dissipation in this tripartite model is constrained by a much tighter, bounded regime determined by the system's configuration space and interaction duration.

As illustrated in Fig.\ref{fig:Thermodynamic bound}(b), the entropy production $\Sigma_\tau$ does not vary arbitrarily above zero but is strictly confined within an upper envelope, establishing an "inaccessible" region of the parametric space. This indicates that the thermodynamic cost is fundamentally tied to the information geometry of the system. To unravel this, we examine the entropy production as a function of the interaction time $\tau$ at a fixed set of parameters $\Lambda = \{\sigma, \omega, \gamma, p_0\}$ in Fig.~\ref{fig:Thermodynamic bound}(c). Remarkably, we find that the finite-time entropy production $\Sigma_\tau$ is strictly constrained by two information-geometric quantities~\citep{SM}:
\begin{equation}
D_{KL}(p^B_0 || p^B_{\tau}) \le \Sigma_\tau \le D_{KL}(p^B_0 || p^B_\infty), \label{bound}
\end{equation}
where $D_{KL}(p||q) = \sum_{i} p_i \ln (p_i/q_i)$ denotes the Kullback-Leibler (KL) divergence, and $p^B_\infty\equiv\lim_{\tau \to \infty} p^B_{\tau}$. The lower bound, $D_{KL}(p^B_0 || p^B_{\tau})$, quantifies the minimum cumulative informational cost required to drive the bit stream from its initial distribution $p_0^B$ to the steady state $p_\tau^B$. Crucially, this inequality rigorously maps to the universal geometric bound of entropy production proven for general relaxation processes~\cite{Naoto2019,shiraishi2023introduction}, confirming that informational friction in our autonomous machine is fundamentally governed by the geometric contraction of the state space. The gap between the actual entropy production $\Sigma_\tau$ and this lower bound originates from the hidden dissipation associated with the transient classical correlations generated between the demon and the bit stream during their finite-time interaction. Physically, the lower bound $D_{KL}(p^B_0 || p^B_{\tau})$ quantifies the minimum thermodynamic cost strictly required to induce the instantaneous informational change in the bits, while the correlation-induced gap embodies the unavoidable penalty of operating at finite speed without quasi-static feedback.

Conversely, the asymptotic upper bound, $D_{KL}(p^B_0 || p^B_\infty)$, dictates the maximal achievable entropy production, $\Sigma_\infty\equiv\lim_{\tau \to \infty} \Sigma_\tau$, for a given parameter set, attained only in the quasi-static limit ($\tau \to \infty$) as correlations vanish. As shown in Fig.~\ref{fig:Thermodynamic bound}(d), $\Delta\Sigma\equiv\Sigma_\infty - \Sigma_\tau \propto e^{2\lambda_3 \tau}$~\cite{SM}, where $\lambda_3$ is the relevant non-zero eigenvalue of the Markov transition matrix $\mathbf{R}$. This explicitly links the macroscopic thermodynamic relaxation of the machine to the spectral gap of its underlying stochastic dynamics. Having established the fundamental information-theoretic constraints governing the system's finite-time relaxation, we now explore how such dynamical bounds dictate the practical performance of the machine.

% \begin{figure}
%     \centering
%     \includegraphics[width=0.5\textwidth]{fig/fig1.png}
%     \caption{(a) Scatter plot of total entropy production $\Sigma_\tau$ versus interaction time $\tau$. The color gradient denotes the initial probability $p_0$ of the incoming bits. (b) For a fixed parameter set $\Lambda$, the finite-time entropy production $\Sigma_{irr}(\tau)$ is tightly bounded between the instantaneous KL divergence $D_{KL}(p_0 || p_\tau)$ and its asymptotic long-time limit $D_{KL}(p_0 || p_\infty)$. (c) Semi-logarithmic plot of the difference $\Sigma_\infty - \Sigma_\tau$ versus $\tau$. The exponential relaxation is governed by $\lambda_3$, the dominant non-zero eigenvalue of the transition rate matrix.}
%     \label{fig:Thermodynamic bound}
% \end{figure}

\textit{Performance qualification of the machine}.---While Eq.~\eqref{bound} dictates the universal bounds of the device, its specific operational mode is determined by the competition between two effective non-equilibrium driven forces, namely, the bit-stream bias $\delta\equiv p_0-p_1$ and the thermal gradient $\epsilon = \tanh[(\beta_c-\beta_h)\Delta E/2]$. The system parameters are recast as $\Lambda=\{\epsilon, \omega, \gamma, \delta\}$. This competition is precisely encoded in the evolution of the outgoing bias, $\delta'(\tau) = \delta - (\delta - \epsilon)\Theta(\tau)$, where the dimensionless function $\Theta(\tau)\in [0,1]$ characterizes the degree of relaxation toward thermal equilibrium and also depends on system parameters $\Lambda$~\cite{SM}. By evaluating the signs of the heat flux $\Delta Q(\tau) = \Delta E (\delta - \epsilon)\Theta(\tau)/2$ and the associated bit-entropy change $\Delta S_B(\tau) = S_B(\delta') - S_B(\delta)$ (with $S_B(\delta)=- \sum_{i=0,1} p_i \ln p_i$), two distinct functional regimes naturally emerge. The machine acts either as a \textit{refrigerator} ($\Delta Q > 0$, $\Delta S_B > 0$) that pumps heat against the thermal gradient by consuming informational order, or as an \textit{eraser} ($\Delta Q < 0$, $\Delta S_B < 0$) that reduces bit entropy via heat dissipation. The power and efficiency of the dual-function machine can be uniformly defined as~\cite{fc2025}
\begin{equation}
P\equiv\frac{\Delta Q^{\frac{1+\theta}{2}}|\Delta S_B|^{\frac{1-\theta}{2}}}{\tau}, \qquad 
\eta\equiv\left(\frac{\Delta \beta\,\Delta Q}{\Delta S_B}\right)^{\theta},
\label{P_E}
\end{equation}
where $\Delta \beta\equiv\beta_c-\beta_h$ and $\theta\equiv {\rm{sign}}({\Delta Q})$. Applying Eqs.~\eqref{bound} and \eqref{Entropyproduction} to there definitions imposes strict thermodynamic bounds on the machine's performance. For instance, the efficiency of the eraser, $\eta_E$, is bounded by
\begin{equation}
\frac{D_{KL}(p^B_0 || p^B_{\tau})}{|\Delta S_B|} < \frac{1-\eta_E}{\eta_E} \le \frac{D_{KL}(p^B_0 || p^B_\infty)}{|\Delta S_B|}. \label{boundeff}
\end{equation}
\begin{figure}   \includegraphics[width=0.5\textwidth]
{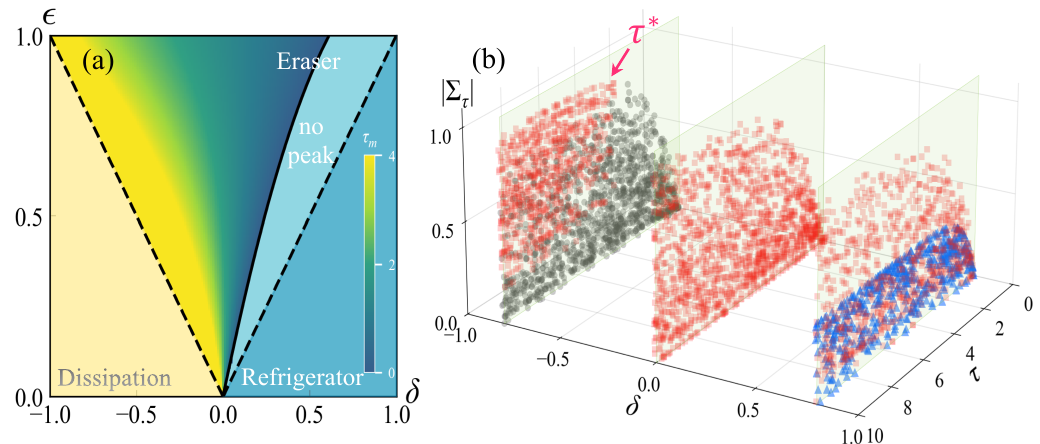}
    \caption{(a) Functional phase diagram in the $(\delta, \epsilon)$ plane at $\omega=0.5$. The color scale maps the optimal time $\tau_m$. Dashed lines ($\epsilon = |\delta|$) delineate the Eraser, Refrigerator, and Dissipative regimes. The solid line bounds the region where an interior power peak exists. (b) Entropy production across the functional regions of the machine for $\delta=-0.8,0,0.8$, sampled over $\omega,\epsilon \in (0,1)$. Functional modes: grey circles (dissipation), red squares (eraser), and blue triangles (refrigerator). }
    \label{fig:taum_Pm_de}
\end{figure}

The functional regimes of the machine, fundamentally dictated by the sign of the heat flux $\mathrm{sign}(\Delta Q) = \mathrm{sign}(\delta - \epsilon)$, are mapped in the phase diagram [Fig.~\ref{fig:taum_Pm_de}(a)] and dynamically corroborated by their finite-time entropy production ($|\Sigma_\tau|$) [Fig.~\ref{fig:taum_Pm_de}(b)]. Distinct colored markers in Fig.~\ref{fig:taum_Pm_de}(b) cleanly partition these operational modes across the parameter space. For a positive initial bias ($\delta > 0$), information resources can dominate ($\delta > \epsilon$) to drive a refrigerator [blue triangles in Fig.~\ref{fig:taum_Pm_de}(b)]. Conversely, if the thermal gradient strongly dominates ($|\epsilon| > |\delta|$), the machine functions as an eraser (red squares). As depicted in the $\delta = 0.8$ slice of Fig.~\ref{fig:taum_Pm_de}(b), while both modes can exist for a positive bias depending on $\epsilon$, they are mutually exclusive. Notably, the refrigeration branch exhibits a distinctively smaller $|\Sigma_\tau|$ (compared to erasure) with a obvious upper bound. This intrinsic cap arises because refrigeration relies on consuming the finite informational order of the incoming bits; since the available negentropy per bit is strictly bounded by $\ln 2$, the maximum dissipation the machine can incur while maintaining its cooling function is fundamentally limited. When the initial bias is perfectly neutral ($\delta = 0$, middle slice), the machine strictly performs pure erasure across all $\tau$.

\begin{figure*}   
    \centering
    \includegraphics[width=1\linewidth]
{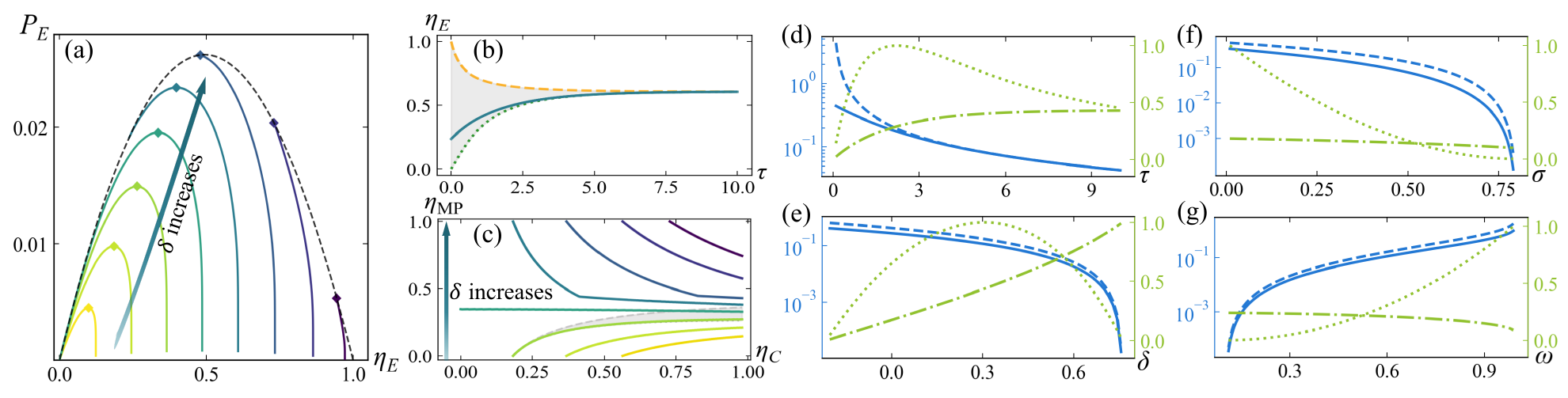}
    \caption{(a) Power-efficiency ($P_E$--$\eta_E$) tradeoff relations for increasing $\delta$ ($\delta=-0.3,-0.2,-0.1,0,0.1,0.2,0.3,0.38$). Diamonds mark the maximum power points. (b) Irreversible entropy production bound for $\eta_E$ as a function of $\tau$, for $\delta=0.1$. The orange dashed line (upper bound) and green dotted line (lower bound) are determined by the Eq.~\eqref{boundeff}. (c) Efficiency at maximum power ${\eta}_{\text{MP}}$, versus $\eta_C$, for the same values of $\delta$ as in panel (a), along with the gray dashed and dotted lines, represent the irreversible entropy production bounds. 
    $\omega=0.5$ and $\epsilon=0.4$ throughout three panels. (d-g) Evolution of thermodynamic quantities as a function of (d) $\tau$, (e) $\delta$, (f) $\sigma$, and (g) $\omega$. In each panel, the remaining parameters are fixed at $\delta=0, \sigma=0.1, \omega=0.8$, and $\tau=1$. The left $y$ axis tracks $P_E(1-\eta_E)/\eta_E$ (solid line) and $D_{KL}(p^B_0 \Vert p^B_\infty)/\tau$ (dashed line), while the right $y$ axis tracks efficiency $\eta_E$ (dash-dotted line) and normalized power $P_E$ (dotted line). 
    }
    \label{fig:eta}
\end{figure*}

The dynamics become particularly rich for a negative initial bias ($\delta < 0$). In the extreme region where $\delta < -\epsilon$ [Dissipation region in Fig.~\ref{fig:taum_Pm_de}(a)], the machine acts purely dissipatively [grey circles in Fig.~\ref{fig:taum_Pm_de}(b)]. However, within the theoretical eraser boundaries ($\epsilon > \delta > -\epsilon$), erasure ($\Delta S_B < 0$) does not occur instantaneously. Instead, the system initially undergoes pure dissipation before functionally transitioning into an eraser. This dynamical crossover is visibly captured in the $\delta = -0.8$ slice of Fig.~\ref{fig:taum_Pm_de}(b) as the sharp boundary $\tau^*$ separating initial dissipation (grey circles) from functional erasure (red squares). Defined by the condition $|\delta'(\tau^*)| = |\delta|$, this transition yields the onset threshold
\begin{equation}
\Theta(\tau^*)=\frac{2\delta}{\delta-\epsilon},
\end{equation}
which precisely quantifies the minimal interaction time required for the thermal gradient to overcome the intrinsic relaxation of the adverse initial bias~\cite{SM}.

\par %Having established the operational boundaries of the eraser, we shift our focus to the finite-time optimization of its performance. Unlike the monotonic decay typical of refrigerator, the erasure power $P_E(\tau)$ exhibits a non-trivial dynamical behavior.
To explore the trade-off between power and efficiency, we investigate the finite-time optimization of the eraser. As illustrated by the parametric $P_E$-$\eta_E$ trajectories in Fig.~\ref{fig:eta}(a), increasing the bit-stream bias $\delta$ alters the machine's operational behavior. The curves transition from exhibiting a power peak ($P_E$ initially rising from a finite value before eventually decaying) to a purely monotonic decay. Because the erasure power approaches a finite value in the short-time limit ($\tau \to 0$) and vanishes when $\tau \to \infty$, the existence of the power peak dictates the optimization strategy. A peak emerges only if the initial rate of entropy reduction outpaces the temporal cost, requiring a positive first-order derivative ($A_1 > 0$) in the short-time expansion $P_E(\tau) = P_E(0) + A_1 \tau + \mathcal{O}(\tau^2)$. This yields the strict criterion
\begin{equation}
\frac{c_2}{\kappa^2}
>
-\frac{\epsilon-\delta}
{2(1-\delta^2)\operatorname{artanh}(\delta)},
\label{eq:bcriterion}
\end{equation}
where $\kappa$ and $c_2$ are coefficients from the expansion $\Theta(\tau)=\kappa\,\tau+c_2\,\tau^2+O(\tau^3)$~\cite{SM}. When this condition is met, $P_E$ attains its maximum at an optimal time $\tau_m$, governed by
\begin{equation}\tau_m(\delta-\epsilon)\frac{\partial \Theta}{\partial \tau}|_{\tau=\tau_m}\operatorname{artanh}(\delta-\delta')=S_B(\delta')-S_B(\delta)\label{eq:taum_equation}.
\end{equation}
By color-mapping $\tau_m$, Fig.~\ref{fig:taum_Pm_de}(a) extends the conventional functional phase diagram of the MQJ model~\cite{2013PRLMQJ} into a quantitative performance optimization landscape. The color gradient explicitly visualizes how $\tau_m$ decreases (transitioning from yellow to dark blue) as $\delta$ grows, reflecting a more immediate power peak in the short-time regime. Consequently, this landscape classifies the eraser into two distinct operational modes. The first mode features a finite-time power peak ($\tau_m > 0$). Remarkably, for $\tau < \tau_m$, this mode operates in a synergistic regime where both power and efficiency increase simultaneously. The second mode corresponds to the distinct ``no-peak" region near the functional boundary. Rather than a performance limitation, this region hosts a fast-running, high-performance eraser: it delivers its absolute maximum power immediately in the short-time limit ($\tau \to 0$) while still maintaining a robust, non-trivial efficiency near the maximum achievable efficiency.  By contrast, in the refrigerator mode, no such peak emerges, and the power–efficiency trade-off remains strictly monotonic, as detailed in ~\cite{SM}.

Figure~\ref{fig:eta}(b) shows that $\eta_E$ is tightly confined by the dissipation bounds derived from Eq.~\eqref{boundeff}. These bounds converge as $\tau$ increases, explicitly linking operational performance to irreversible entropy production. To evaluate the thermodynamic quality of the eraser, we examine the efficiency at maximum power (EMP), $\eta_{\text{MP}}$ [Fig.~\ref{fig:eta}(c)]. The EMP curves originate from an operational cutoff $\eta_C^{\min} = \operatorname{artanh}(|\delta|) / \operatorname{artanh}(\omega)$, strictly dictated by the functional boundary $|\epsilon| > |\delta|$.Despite this symmetric onset regarding the magnitude $|\delta|$, the EMP behavior is highly asymmetric with respect to the bias direction. For an adverse bias ($\delta \le 0$), $\eta_{\text{MP}}$ decreases monotonically with $\eta_C$, suppressed by the substantial dissipation required to overcome the initial disorder. Conversely, for a cooperative bias ($\delta > 0$), $\eta_{\text{MP}}$ approaches unity near the functional onset ($\eta_C \to \eta_{C}^{\min}$), indicating near-reversible operation even at maximum power. Finally, substituting the optimal time $\tau_m$ [Eq.~\eqref{eq:taum_equation}] into Eq.~\eqref{boundeff} yields rigorous irreversible bounds on $\eta_{\text{MP}}$, effectively partitioning the allowable performance space (e.g., the shaded region in Fig.~\ref{fig:eta}(c) for $\delta = -0.1$).

% d   plot_vs_tau(delta=0, sigma=0.1, omega=0.8, gamma=1.0)
% e   plot_vs_delta(tau=1, sigma=0.1, omega=0.8, gamma=1.0)
% f   plot_vs_sigma(delta=0, tau=1, omega=0.8, gamma=1.0)
% g   plot_vs_omega(delta=0, sigma=0.1, tau=1, gamma=1.0)

Moreover, combining Eqs.~\eqref{P_E} and \eqref{boundeff} yields a joint upper bound constraining the efficiency and power
\begin{equation}\frac{P_E(1-\eta_E)}{\eta_E} \le \frac{D_{KL}(p^B_0 || p^B_\infty)}{\tau}.
\end{equation}
which is plotted as blue dashed curves in Figs.~\ref{fig:eta}(d)-(g) in comparison with the exact performance metric (blue solid curves). The bound strictly confines the exact results, becoming progressively tighter as the interaction time $\tau$ increases [Fig.~\ref{fig:eta}(d)]. Remarkably, it also maintains exceptional tightness across broad parametric sweeps of $\delta$, $\sigma$, and $\omega$ [Figs.~\ref{fig:eta}(e-g)]. Crucially, rather than enforcing a strict trade-off, the functional structure of this bound permits the simultaneous enhancement of both $P_E$ and $\eta_E$. As clearly seen in the short-time region ($\tau\lesssim2$) in [Fig.~\ref{fig:eta}(d)] and the $\delta\lesssim0.3$ region in [Fig.~\ref{fig:eta}(e)], the machine enters a unique synergistic window where power (green dotted curves) and efficiency (green dash-dotted curves) robustly increase in tandem. Furthermore, exploring the parameter space reveals a highly effective optimization strategy: as shown in Figs.~\ref{fig:eta}(f) and (g), by decreasing $\sigma$ or increasing $\omega$, the erasure power can be significantly amplified while sustaining a finite and remarkably stable efficiency.

\textit{Conclusion and Outlook}.---In summary, the finite-time thermodynamic limits of the autonomous information machine are fundamentally dictated by the temporal lag between the demon's finite relaxation and the incoming bit stream. By restricting energy fluxes through internal thermal transition rates, this intrinsic mechanism prevents operational power from diverging and dictates the highly asymmetric throughput-efficiency trade-off in the high-frequency limit. Crucially, our study exposes the profound consequences of removing the idealized external agent. We reveal that unavoidable backaction within the coupled subsystems imposes strict upper bounds on the maximum achievable erasure and cooling rates. These findings establish a comprehensive theoretical framework for dynamical optimization, demonstrating that non-equilibrium limits become fundamentally more restrictive when systems operate entirely free of external control.

Looking forward, our findings pave the way for broader investigations in finite-time information thermodynamics. The established bounds provide a foundational starting point for thermodynamic geometry analysis to systematically uncover dissipation-minimizing operational pathways~\cite{Ito2018,Scandi2022,2022Ma}. Moreover, extending this framework to more complex environments, such as information reservoirs with multi-bit temporal correlations~\citep{Lorenzo2015}, or into the quantum regime where bit and demon coherence come into play~\cite{Chapman2015, Lostaglio2015, Goold2015,Snchez2019,Miller2020}, raises compelling open questions. Specifically, it remains to be explored whether such classical correlations or quantum effects act as thermodynamic burdens, or if they can be actively harvested as non-equilibrium resources to surpass current classical speed limits~\citep{Funo2013,Still2020}. Ultimately, bridging these theoretical advancements will provide crucial design principles for synthetic molecular machines and biological active systems~\cite{2024Chen,Leighton2025}, where the optimal synchronization of thermodynamic and informational fluxes under strict temporal constraints is paramount.

\textit{Acknowledgment---.} We thank Y. Q. Lin and X. H. Zhao for fruitful discussion. This work is supported by the National Natural Science Foundation under Grant No. 12305037.
%Although each interaction is a small step, this is compensated by the large number of cycles, whose cumulative effect yields overall optimal performance.This reveals a distinct fast-autonomous advantage in the regime of small Carnot efficiency ($\eta_c$): operating at high frequencies enables the MQJ machine to approach near-Carnot performance ($\eta_m \approx \eta_c$) and achieve substantial cumulative erasure. 

%Remarkably, the performance landscape in Fig.~\ref{fig:taum_Pm_de}(b) reveals a unique cooperative regime for $\tau < \tau_m$, the system enters a cooperative regime where both $P_E$ and the erasure efficiency $\eta_E$ increase monotonically with $\tau$. In this window, the device circumvents the conventional power-efficiency trade-off, as moving toward the "sweet spot" $\tau_m$ simultaneously improves both the rate and the quality of information processing. Furthermore, as $\delta$ increases, the optimal $(P_m, \eta_m)$ first rise with reduced initial entropy easing erasure, then fall as $\delta \to \epsilon$ and the thermodynamic drive vanishes. This suggests that by judiciously tuning the bit-stream bias $\delta$ and the interaction duration, one can synchronize the informational and thermal currents to stabilize an extended cooperative window, thereby achieving a global optimization of high power and high efficiency. 

\bibliography{refs}

@article{parrondo_thermodynamics_2015,
	title = {Thermodynamics of information},
	volume = {11},
	issn = {1745-2473, 1745-2481},
	url = {https://www.nature.com/articles/nphys3230},
	doi = {10.1038/nphys3230},
	number = {2},
	journal = {Nat. Phys.},
	author = {Parrondo, Juan M. R. and Horowitz, Jordan M. and Sagawa, Takahiro},
	month = feb,
	year = {2015},
	pages = {131--139},
}

@article{Gour2015,
  title = {The resource theory of informational nonequilibrium in thermodynamics},
  volume = {583},
  ISSN = {0370-1573},
  url = {http://dx.doi.org/10.1016/j.physrep.2015.04.003},
  DOI = {10.1016/j.physrep.2015.04.003},
  journal = {Phys. Rep.},
  publisher = {Elsevier BV},
  author = {Gour,  Gilad and M\"{u}ller,  Markus P. and Narasimhachar,  Varun and Spekkens,  Robert W. and Yunger Halpern,  Nicole},
  year = {2015},
  month = jul,
  pages = {1–58}
}

@article{Sagawa2012,
  title = {Fluctuation Theorem with Information Exchange: Role of Correlations in Stochastic Thermodynamics},
  author = {Sagawa, Takahiro and Ueda, Masahito},
  journal = {Phys. Rev. Lett.},
  volume = {109},
  issue = {18},
  pages = {180602},
  numpages = {5},
  year = {2012},
  month = {Nov},
  publisher = {American Physical Society},
  doi = {10.1103/PhysRevLett.109.180602},
  url = {https://link.aps.org/doi/10.1103/PhysRevLett.109.180602}
}

@article{Horowitz2014,
  title = {Thermodynamics with Continuous Information Flow},
  author = {Horowitz, Jordan M. and Esposito, Massimiliano},
  journal = {Phys. Rev. X},
  volume = {4},
  issue = {3},
  pages = {031015},
  numpages = {11},
  year = {2014},
  month = {Jul},
  publisher = {American Physical Society},
  doi = {10.1103/PhysRevX.4.031015},
  url = {https://link.aps.org/doi/10.1103/PhysRevX.4.031015}
}

@article{Wolpert2019,
  title = {The stochastic thermodynamics of computation},
  volume = {52},
  ISSN = {1751-8121},
  url = {http://dx.doi.org/10.1088/1751-8121/ab0850},
  DOI = {10.1088/1751-8121/ab0850},
  number = {19},
  journal = {‌J. Phys. A: Math. Theor.},
  publisher = {IOP Publishing},
  author = {Wolpert,  David H},
  year = {2019},
  month = apr,
  pages = {193001}
}

@article{Landauer1961,
  title = {Irreversibility and Heat Generation in the Computing Process},
  volume = {5},
  ISSN = {0018-8646},
  url = {http://dx.doi.org/10.1147/rd.53.0183},
  DOI = {10.1147/rd.53.0183},
  number = {3},
  journal = {IBM J. Res. Dev.},
  publisher = {IBM},
  author = {Landauer,  R.},
  year = {1961},
  month = jul,
  pages = {183–191}
}

@article{Bennett1982,
  title = {The thermodynamics of computation—a review},
  volume = {21},
  ISSN = {1572-9575},
  url = {http://dx.doi.org/10.1007/bf02084158},
  DOI = {10.1007/bf02084158},
  number = {12},
  journal = {Int. J. Theor. Phys.},
  publisher = {Springer Science and Business Media LLC},
  author = {Bennett,  Charles H.},
  year = {1982},
  month = dec,
  pages = {905–940}
}

@article{Brut2012,
  title = {Experimental verification of Landauer’s principle linking information and thermodynamics},
  volume = {483},
  ISSN = {1476-4687},
  url = {http://dx.doi.org/10.1038/nature10872},
  DOI = {10.1038/nature10872},
  number = {7388},
  journal = {Nature},
  publisher = {Springer Science and Business Media LLC},
  author = {Bérut,  Antoine and Arakelyan,  Artak and Petrosyan,  Artyom and Ciliberto,  Sergio and Dillenschneider,  Raoul and Lutz,  Eric},
  year = {2012},
  month = mar,
  pages = {187–189}
}

@article{Proesmans2020,
  title = {Finite-Time Landauer Principle},
  author = {Proesmans, Karel and Ehrich, Jannik and Bechhoefer, John},
  journal = {Phys. Rev. Lett.},
  volume = {125},
  issue = {10},
  pages = {100602},
  numpages = {6},
  year = {2020},
  month = {Sep},
  publisher = {American Physical Society},
  doi = {10.1103/PhysRevLett.125.100602},
  url = {https://link.aps.org/doi/10.1103/PhysRevLett.125.100602}
}

@article{Zhen2021,
  title = {Universal Bound on Energy Cost of Bit Reset in Finite Time},
  author = {Zhen, Yi-Zheng and Egloff, Dario and Modi, Kavan and Dahlsten, Oscar},
  journal = {Phys. Rev. Lett.},
  volume = {127},
  issue = {19},
  pages = {190602},
  numpages = {7},
  year = {2021},
  month = {Nov},
  publisher = {American Physical Society},
  doi = {10.1103/PhysRevLett.127.190602},
  url = {https://link.aps.org/doi/10.1103/PhysRevLett.127.190602}
}

@article{2024Zhou,
  title = {Finite-time optimization of a quantum Szilard heat engine},
  author = {Zhou, Tan-Ji and Ma, Yu-Han and Sun, C. P.},
  journal = {Phys. Rev. Res.},
  volume = {6},
  issue = {4},
  pages = {043001},
  numpages = {12},
  year = {2024},
  month = {Oct},
  publisher = {American Physical Society},
  doi = {10.1103/PhysRevResearch.6.043001},
  url = {https://link.aps.org/doi/10.1103/PhysRevResearch.6.043001}
}

@article{2022Ma,
  title = {Minimal energy cost to initialize a bit with tolerable error},
  author = {Ma, Yu-Han and Chen, Jin-Fu and Sun, C. P. and Dong, Hui},
  journal = {Phys. Rev. E},
  volume = {106},
  issue = {3},
  pages = {034112},
  numpages = {8},
  year = {2022},
  month = {Sep},
  publisher = {American Physical Society},
  doi = {10.1103/PhysRevE.106.034112},
  url = {https://link.aps.org/doi/10.1103/PhysRevE.106.034112}
}

@article{Dago2022,
  title = {Dynamics of Information Erasure and Extension of Landauer's Bound to Fast Processes},
  author = {Dago, Salamb\^o and Bellon, Ludovic},
  journal = {Phys. Rev. Lett.},
  volume = {128},
  issue = {7},
  pages = {070604},
  numpages = {6},
  year = {2022},
  month = {Feb},
  publisher = {American Physical Society},
  doi = {10.1103/PhysRevLett.128.070604},
  url = {https://link.aps.org/doi/10.1103/PhysRevLett.128.070604}
}

@article{Lee2022,
  title = {Speed Limit for a Highly Irreversible Process and Tight Finite-Time Landauer's Bound},
  author = {Lee, Jae Sung and Lee, Sangyun and Kwon, Hyukjoon and Park, Hyunggyu},
  journal = {Phys. Rev. Lett.},
  volume = {129},
  issue = {12},
  pages = {120603},
  numpages = {7},
  year = {2022},
  month = {Sep},
  publisher = {American Physical Society},
  doi = {10.1103/PhysRevLett.129.120603},
  url = {https://link.aps.org/doi/10.1103/PhysRevLett.129.120603}
}

@article{Quan2006,
  title = {Maxwell's Demon Assisted Thermodynamic Cycle in Superconducting Quantum Circuits},
  author = {Quan, H. T. and Wang, Y. D. and Liu, Yu-xi and Sun, C. P. and Nori, Franco},
  journal = {Phys. Rev. Lett.},
  volume = {97},
  issue = {18},
  pages = {180402},
  numpages = {4},
  year = {2006},
  month = {Oct},
  publisher = {American Physical Society},
  doi = {10.1103/PhysRevLett.97.180402},
  url = {https://link.aps.org/doi/10.1103/PhysRevLett.97.180402}
}

@article{Toyabe2010,
  title = {Experimental demonstration of information-to-energy conversion and validation of the generalized Jarzynski equality},
  volume = {6},
  ISSN = {1745-2481},
  url = {http://dx.doi.org/10.1038/nphys1821},
  DOI = {10.1038/nphys1821},
  number = {12},
  journal = {Nat. Phys.},
  publisher = {Springer Science and Business Media LLC},
  author = {Toyabe,  Shoichi and Sagawa,  Takahiro and Ueda,  Masahito and Muneyuki,  Eiro and Sano,  Masaki},
  year = {2010},
  month = nov,
  pages = {988–992}
}

@article{Mandal2012,
  title = {Work and information processing in a solvable model of Maxwell’s demon},
  volume = {109},
  ISSN = {1091-6490},
  url = {http://dx.doi.org/10.1073/pnas.1204263109},
  DOI = {10.1073/pnas.1204263109},
  number = {29},
  journal = {Proc. Natl. Acad. Sci. USA},
  publisher = {Proceedings of the National Academy of Sciences},
  author = {Mandal,  Dibyendu and Jarzynski,  Christopher},
  year = {2012},
  month = jul,
  pages = {11641–11645}
}

@article{Fadler2023,
  title = {Efficiency at Maximum Power of a Carnot Quantum Information Engine},
  author = {Fadler, Paul and Friedenberger, Alexander and Lutz, Eric},
  journal = {Phys. Rev. Lett.},
  volume = {130},
  issue = {24},
  pages = {240401},
  numpages = {7},
  year = {2023},
  month = {Jun},
  publisher = {American Physical Society},
  doi = {10.1103/PhysRevLett.130.240401},
  url = {https://link.aps.org/doi/10.1103/PhysRevLett.130.240401}
}

@article{Cai2012,
  title = {Multiparticle quantum Szilard engine with optimal cycles assisted by a Maxwell's demon},
  author = {Cai, C. Y. and Dong, H. and Sun, C. P.},
  journal = {Phys. Rev. E},
  volume = {85},
  issue = {3},
  pages = {031114},
  numpages = {12},
  year = {2012},
  month = {Mar},
  publisher = {American Physical Society},
  doi = {10.1103/PhysRevE.85.031114},
  url = {https://link.aps.org/doi/10.1103/PhysRevE.85.031114}
}

@article{Browne2014,
  title = {Guaranteed Energy-Efficient Bit Reset in Finite Time},
  author = {Browne, Cormac and Garner, Andrew J. P. and Dahlsten, Oscar C. O. and Vedral, Vlatko},
  journal = {Phys. Rev. Lett.},
  volume = {113},
  issue = {10},
  pages = {100603},
  numpages = {5},
  year = {2014},
  month = {Sep},
  publisher = {American Physical Society},
  doi = {10.1103/PhysRevLett.113.100603},
  url = {https://link.aps.org/doi/10.1103/PhysRevLett.113.100603}
}

@article{Danageozian2022,
  title = {Thermodynamic Constraints on Quantum Information Gain and Error Correction: A Triple Trade-Off},
  author = {Danageozian, Arshag and Wilde, Mark M. and Buscemi, Francesco},
  journal = {PRX Quantum},
  volume = {3},
  issue = {2},
  pages = {020318},
  numpages = {21},
  year = {2022},
  month = {Apr},
  publisher = {American Physical Society},
  doi = {10.1103/PRXQuantum.3.020318},
  url = {https://link.aps.org/doi/10.1103/PRXQuantum.3.020318}
}

@article{Scandi2022,
  title = {Minimally Dissipative Information Erasure in a Quantum Dot via Thermodynamic Length},
  author = {Scandi, Matteo and Barker, David and Lehmann, Sebastian and Dick, Kimberly A. and Maisi, Ville F. and Perarnau-Llobet, Mart\'{\i}},
  journal = {Phys. Rev. Lett.},
  volume = {129},
  issue = {27},
  pages = {270601},
  numpages = {6},
  year = {2022},
  month = {Dec},
  publisher = {American Physical Society},
  doi = {10.1103/PhysRevLett.129.270601},
  url = {https://link.aps.org/doi/10.1103/PhysRevLett.129.270601}
}

@article{Erdman2025,
  title = {Artificially intelligent Maxwell’s demon for optimal control of open quantum systems},
  volume = {10},
  ISSN = {2058-9565},
  url = {http://dx.doi.org/10.1088/2058-9565/adbccf},
  DOI = {10.1088/2058-9565/adbccf},
  number = {2},
  journal = {Quantum Sci. Technol.},
  publisher = {IOP Publishing},
  author = {Erdman,  Paolo A and Czupryniak,  Robert and Bhandari,  Bibek and Jordan,  Andrew N and Noé,  Frank and Eisert,  Jens and Guarnieri,  Giacomo},
  year = {2025},
  month = mar,
  pages = {025047}
}

@article{McGrath2017,
  title = {Biochemical Machines for the Interconversion of Mutual Information and Work},
  author = {McGrath, Thomas and Jones, Nick S. and ten Wolde, Pieter Rein and Ouldridge, Thomas E.},
  journal = {Phys. Rev. Lett.},
  volume = {118},
  issue = {2},
  pages = {028101},
  numpages = {5},
  year = {2017},
  month = {Jan},
  publisher = {American Physical Society},
  doi = {10.1103/PhysRevLett.118.028101},
  url = {https://link.aps.org/doi/10.1103/PhysRevLett.118.028101}
}

@article{Amano2022,
  title = {Insights from an information thermodynamics analysis of a synthetic molecular motor},
  volume = {14},
  ISSN = {1755-4349},
  url = {http://dx.doi.org/10.1038/s41557-022-00899-z},
  DOI = {10.1038/s41557-022-00899-z},
  number = {5},
  journal = {Nat. Chem.},
  publisher = {Springer Science and Business Media LLC},
  author = {Amano,  Shuntaro and Esposito,  Massimiliano and Kreidt,  Elisabeth and Leigh,  David A. and Penocchio,  Emanuele and Roberts,  Benjamin M. W.},
  year = {2022},
  month = mar,
  pages = {530–537}
}

@article{2025Jarzynskia,
	title = {Fluctuation theorems for autonomous work},
	volume = {122},
	issn = {0027-8424, 1091-6490},
	url = {https://pnas.org/doi/10.1073/pnas.2524775122},
	doi = {10.1073/pnas.2524775122},
	number = {51},
	journal = {Proc. Natl. Acad. Sci. USA},
	author = {Jarzynski, Christopher and Deffner, Sebastian and Rahav, Saar},
	month = dec,
	year = {2025},
	pages = {e2524775122},
}

@article{2013PRLMQJ,
	title = {Maxwell’s {Refrigerator}: {An} {Exactly} {Solvable} {Model}},
	volume = {111},
	copyright = {http://link.aps.org/licenses/aps-default-license},
	issn = {0031-9007, 1079-7114},
	shorttitle = {Maxwell’s {Refrigerator}},
	url = {https://link.aps.org/doi/10.1103/PhysRevLett.111.030602},
	doi = {10.1103/PhysRevLett.111.030602},
	number = {3},
	journal = {Phys. Rev. Lett.},
	author = {Mandal, Dibyendu and Quan, H. T. and Jarzynski, Christopher},
	month = jul,
	year = {2013},
	pages = {030602},
}

@article{Ito2018,
  title = {Stochastic Thermodynamic Interpretation of Information Geometry},
  author = {Ito, Sosuke},
  journal = {Phys. Rev. Lett.},
  volume = {121},
  issue = {3},
  pages = {030605},
  numpages = {7},
  year = {2018},
  month = {Jul},
  doi = {10.1103/PhysRevLett.121.030605},
  url = {https://link.aps.org/doi/10.1103/PhysRevLett.121.030605}
}

@article{Lorenzo2015,
  title = {Landauer's Principle in Multipartite Open Quantum System Dynamics},
  author = {Lorenzo, S. and McCloskey, R. and Ciccarello, F. and Paternostro, M. and Palma, G. M.},
  journal = {Phys. Rev. Lett.},
  volume = {115},
  issue = {12},
  pages = {120403},
  numpages = {5},
  year = {2015},
  month = {Sep},
  publisher = {American Physical Society},
  doi = {10.1103/PhysRevLett.115.120403},
  url = {https://link.aps.org/doi/10.1103/PhysRevLett.115.120403}
}

@article{Lostaglio2015,
	title = {Description of quantum coherence in thermodynamic processes requires constraints beyond free energy},
	volume = {6},
	issn = {2041-1723},
	url = {https://www.nature.com/articles/ncomms7383},
	doi = {10.1038/ncomms7383},
	number = {1},
	journal = {Nat. Commu.},
	author = {Lostaglio, Matteo and Jennings, David and Rudolph, Terry},
	month = mar,
	year = {2015},
	pages = {6383},
}

@article{Goold2015,
  title = {Nonequilibrium Quantum Landauer Principle},
  author = {Goold, John and Paternostro, Mauro and Modi, Kavan},
  journal = {Phys. Rev. Lett.},
  volume = {114},
  issue = {6},
  pages = {060602},
  numpages = {5},
  year = {2015},
  month = {Feb},
  publisher = {American Physical Society},
  doi = {10.1103/PhysRevLett.114.060602},
  url = {https://link.aps.org/doi/10.1103/PhysRevLett.114.060602}
}

@article{Snchez2019,
  title = {Autonomous conversion of information to work in quantum dots},
  author = {S\'anchez, Rafael and Samuelsson, Peter and Potts, Patrick P.},
  journal = {Phys. Rev. Res.},
  volume = {1},
  issue = {3},
  pages = {033066},
  numpages = {18},
  year = {2019},
  month = {Oct},
  publisher = {American Physical Society},
  doi = {10.1103/PhysRevResearch.1.033066},
  url = {https://link.aps.org/doi/10.1103/PhysRevResearch.1.033066}
}

@article{Miller2020,
  title = {Quantum Fluctuations Hinder Finite-Time Information Erasure near the Landauer Limit},
  author = {Miller, Harry J. D. and Guarnieri, Giacomo and Mitchison, Mark T. and Goold, John},
  journal = {Phys. Rev. Lett.},
  volume = {125},
  issue = {16},
  pages = {160602},
  numpages = {6},
  year = {2020},
  month = {Oct},
  publisher = {American Physical Society},
  doi = {10.1103/PhysRevLett.125.160602},
  url = {https://link.aps.org/doi/10.1103/PhysRevLett.125.160602}
}

@article{kosloff2014quantum,
  title={Quantum heat engines and refrigerators: Continuous devices},
  author={Kosloff, Ronnie and Levy, Amikam},
  journal={Annual review of physical chemistry},
  volume={65},
  number={1},
  pages={365--393},
  year={2014},
  publisher={Annual Reviews},
doi={10.1146/annurev-physchem-040513-103724}
}

@Article{Kosloff2013,
AUTHOR = {Kosloff, Ronnie},
TITLE = {Quantum Thermodynamics: A Dynamical Viewpoint},
JOURNAL = {Entropy},
VOLUME = {15},
YEAR = {2013},
NUMBER = {6},
PAGES = {2100--2128},
URL = {https://www.mdpi.com/1099-4300/15/6/2100},
ISSN = {1099-4300},
DOI = {10.3390/e15062100}
}

@article{shiraishi2023introduction,
  title={An introduction to stochastic thermodynamics},
  author={Shiraishi, Naoto},
  journal={Fundamental Theories of Physics. Springer, Singapore},
  year={2023},
  publisher={Springer},
  doi={https://doi.org/10.1007/978-981-19-8186-9}
}

@article{Naoto2019,
  title = {Information-Theoretical Bound of the Irreversibility in Thermal Relaxation Processes},
  author = {Shiraishi, Naoto and Saito, Keiji},
  journal = {Phys. Rev. Lett.},
  volume = {123},
  issue = {11},
  pages = {110603},
  numpages = {6},
  year = {2019},
  month = {Sep},
  publisher = {American Physical Society},
  doi = {10.1103/PhysRevLett.123.110603},
  url = {https://link.aps.org/doi/10.1103/PhysRevLett.123.110603}
}

@Article{Bjarne2022,
AUTHOR = {Andresen, Bjarne and Salamon, Peter},
TITLE = {Future Perspectives of Finite-Time Thermodynamics},
JOURNAL = {Entropy},
VOLUME = {24},
YEAR = {2022},
NUMBER = {5},
ARTICLE-NUMBER = {690},
URL = {https://www.mdpi.com/1099-4300/24/5/690},
PubMedID = {35626573},
ISSN = {1099-4300},
DOI = {10.3390/e24050690}
}

@article{Funo2013,
  title = {Thermodynamic work gain from entanglement},
  author = {Funo, Ken and Watanabe, Yu and Ueda, Masahito},
  journal = {Phys. Rev. A},
  volume = {88},
  issue = {5},
  pages = {052319},
  numpages = {7},
  year = {2013},
  month = {Nov},
  publisher = {American Physical Society},
  doi = {10.1103/PhysRevA.88.052319},
  url = {https://link.aps.org/doi/10.1103/PhysRevA.88.052319}
}

@article{Still2020,
  title = {Thermodynamic Cost and Benefit of Memory},
  author = {Still, Susanne},
  journal = {Phys. Rev. Lett.},
  volume = {124},
  issue = {5},
  pages = {050601},
  numpages = {6},
  year = {2020},
  month = {Feb},
  publisher = {American Physical Society},
  doi = {10.1103/PhysRevLett.124.050601},
  url = {https://link.aps.org/doi/10.1103/PhysRevLett.124.050601}
}

@article{2024Chen,
	title = {Efficiency of an autonomous, dynamic information engine operating on a single active particle},
	volume = {110},
	issn = {2470-0045, 2470-0053},
	url = {https://link.aps.org/doi/10.1103/PhysRevE.110.014602},
	doi = {10.1103/PhysRevE.110.014602},
	number = {1},
	journal = {Phys. Rev. E},
	author = {Cocconi, Luca and Chen, Letian},
	month = jul,
	year = {2024},
	pages = {014602},
}

@article{Leighton2025,
  title = {Flow of Energy and Information in Molecular Machines},
  volume = {76},
  ISSN = {1545-1593},
  url = {http://dx.doi.org/10.1146/annurev-physchem-082423-030023},
  DOI = {10.1146/annurev-physchem-082423-030023},
  number = {1},
  journal = {Annu. Rev. Phys. Chem.},
  publisher = {Annual Reviews},
  author = {Leighton,  Matthew P. and Sivak,  David A.},
  year = {2025},
  month = apr,
  pages = {379–403}
}

@misc{SM,

note = {See Supplemental Materials at [URL] for detailed derivations and supporting informations}
}

@article{fc2025,
  title = {Unified approach to power-efficiency trade-off relations of generic thermal machines},
  author = {Ma, Yu-Han and Fu, Cong},
  journal = {Phys. Rev. E},
  volume = {112},
  issue = {5},
  pages = {054130},
  numpages = {10},
  year = {2025},
  month = {Nov},
  publisher = {American Physical Society},
  doi = {10.1103/bvlw-rvvv},
  url = {https://link.aps.org/doi/10.1103/bvlw-rvvv}
}

@article{qiu20roadmap,
  title={Roadmap on thermodynamics and thermal metamaterials},
  author={Qiu, Yuguang and Nomura, Masahiro and Zhang, Zhongwei and others},
  journal={Front. Phys.},
  volume={20},
  number={6},
  pages={065500},
  doi={10.15302/frontphys.2025.065500}
}

@article{2022Van,
  title = {Finite-Time Quantum Landauer Principle and Quantum Coherence},
  author = {Van Vu, Tan and Saito, Keiji},
  journal = {Phys. Rev. Lett.},
  volume = {128},
  issue = {1},
  pages = {010602},
  numpages = {7},
  year = {2022},
  month = {Jan},
  publisher = {American Physical Society},
  doi = {10.1103/PhysRevLett.128.010602},
  url = {https://link.aps.org/doi/10.1103/PhysRevLett.128.010602}
}

@article{Barato2014,
  title = {Stochastic thermodynamics with information reservoirs},
  author = {Barato, Andre C. and Seifert, Udo},
  journal = {Phys. Rev. E},
  volume = {90},
  issue = {4},
  pages = {042150},
  numpages = {11},
  year = {2014},
  month = {Oct},
  publisher = {American Physical Society},
  doi = {10.1103/PhysRevE.90.042150},
  url = {https://link.aps.org/doi/10.1103/PhysRevE.90.042150}
}

@article{Zulkowski2014,
  title = {Optimal finite-time erasure of a classical bit},
  author = {Zulkowski, Patrick R. and DeWeese, Michael R.},
  journal = {Phys. Rev. E},
  volume = {89},
  issue = {5},
  pages = {052140},
  numpages = {9},
  year = {2014},
  month = {May},
  publisher = {American Physical Society},
  doi = {10.1103/PhysRevE.89.052140},
  url = {https://link.aps.org/doi/10.1103/PhysRevE.89.052140}
}

@article{cao2015,
  title = {Thermodynamics of information processing based on enzyme kinetics: An exactly solvable model of an information pump},
  author = {Cao, Yuansheng and Gong, Zongping and Quan, H. T.},
  journal = {Phys. Rev. E},
  volume = {91},
  issue = {6},
  pages = {062117},
  numpages = {11},
  year = {2015},
  month = {Jun},
  publisher = {American Physical Society},
  doi = {10.1103/PhysRevE.91.062117},
  url = {https://link.aps.org/doi/10.1103/PhysRevE.91.062117}
}

@article{Bao2023,
  title = {Designing autonomous Maxwell's demon via stochastic resetting},
  author = {Bao, Ruicheng and Cao, Zhiyu and Zheng, Jiming and Hou, Zhonghuai},
  journal = {Phys. Rev. Res.},
  volume = {5},
  issue = {4},
  pages = {043066},
  numpages = {17},
  year = {2023},
  month = {Oct},
  publisher = {American Physical Society},
  doi = {10.1103/PhysRevResearch.5.043066},
  url = {https://link.aps.org/doi/10.1103/PhysRevResearch.5.043066}
}

@article{Cao2023Chem,
author = {Cao, Zhiyu and Bao, Ruicheng and Zheng, Jiming and Hou, Zhonghuai},
title = {Fast Functionalization with High Performance in the Autonomous Information Engine},
journal = {J. Phys. Chem. Lett.},
volume = {14},
number = {1},
pages = {66-72},
year = {2023},
doi = {10.1021/acs.jpclett.2c03335},

}

@article{Chapman2015,
  title = {How an autonomous quantum Maxwell demon can harness correlated information},
  author = {Chapman, Adrian and Miyake, Akimasa},
  journal = {Phys. Rev. E},
  volume = {92},
  issue = {6},
  pages = {062125},
  numpages = {12},
  year = {2015},
  month = {Dec},
  publisher = {American Physical Society},
  doi = {10.1103/PhysRevE.92.062125},
  url = {https://link.aps.org/doi/10.1103/PhysRevE.92.062125}
}

@article{aguilar2025power,
  title={Power-efficiency-stability trade-off in quantum information engines},
  author={Aguilar, Milton and {\"U}nal, C{\"u}neyt and Lutz, Eric},
  journal={arXiv preprint arXiv:2509.04901},
  year={2025},
  url = {https://arxiv.org/abs/2509.04901},
  publisher = {arXiv},
}

@article{Tushar2021,
    author = {Tushar K. Saha  and Joseph N. E. Lucero  and Jannik Ehrich  and David A. Sivak  and John Bechhoefer },
    title = {Maximizing power and velocity of an information engine},
    journal = {Proc. Natl. Acad. Sci. USA},
    volume = {118},
    number = {20},
    pages = {e2023356118},
    year = {2021},
    doi = {10.1073/pnas.2023356118},
    URL = {https://www.pnas.org/doi/abs/10.1073/pnas.2023356118}
}

\end{document}

\title{Supplemental Materials for: \\ ``Finite-Time Thermodynamics of an Autonomous Information Machine''}       
\author{Wanyan Chen}
\affiliation{School of Physics and Astronomy, Beijing Normal University, Beijing, 100875, China}
\affiliation{Key Laboratory of Multiscale Spin Physics (Ministry of Education), Beijing Normal University,Beijing 100875, China}

\author{Miao Chen}
\affiliation{School of Physics and Astronomy, Beijing Normal University, Beijing, 100875, China}
\affiliation{Key Laboratory of Multiscale Spin Physics (Ministry of Education), Beijing Normal University,Beijing 100875, China}

\author{Yu-Han Ma}
\email{yhma@bnu.edu.cn}
\affiliation{School of Physics and Astronomy, Beijing Normal University, Beijing, 100875, China}
\affiliation{Key Laboratory of Multiscale Spin Physics (Ministry of Education), Beijing Normal University,Beijing 100875, China}

\maketitle
%\end{CJK*}
\onecolumngrid
\begin{spacing}{1.7} 
\textbf{This document is devoted to providing the detailed derivations and the supporting discussions to the main text of the Letter. The contents of the Supplemental Materials are listed as follows}

\tableofcontents

\end{spacing}
\newpage

\setlength{\parskip}{1em}

\section{Information-geometric Bounds of $\Sigma_\tau$}

In this section, we derive the upper and lower bounds for the total entropy production $\Sigma_\tau$ based on Kullback-Leibler (KL) divergence in [Eq.~(6)] of the main text. Furthermore, we examine the asymptotic behavior of $\Sigma_\tau$ towards $\Sigma_\infty$. 

To establish these bounds, we first characterize the evolution of the bit distribution, which defines the geometric structure of the KL inequalities. The bit stream evolution is described by $\delta$, $\delta'$, and the stationary limit $\delta_\infty$. Following the framework and derivations in the supplemental material of Ref.~\cite{2013PRLMQJ}, where $f(\delta') > f(\delta)$ with $f(\delta) = K\delta + S(\delta)$ a strictly concave function maximized at $\epsilon=\delta_\infty$, and $\operatorname{sign}(\delta - \delta') = \operatorname{sign}(\delta - \epsilon)$, it follows that $\delta'$ lies between $\delta$ and $\delta_\infty$. Equivalently, $p_0'$ lies between $p_0$ and $p_\infty$.

With the above constraints on the evolution of the distribution, we now derive the upper and lower bounds for $\Sigma_\tau$. First, $\Sigma_\tau$ can be rewritten as
\begin{equation}
    \Sigma_\tau  = D_{KL}(p_0^B || p_\infty^B) - D_{KL}(p_\tau^B || p_\infty^B). 
    \label{eq:Sigma}
\end{equation}
Since $D_{KL}(p||q)$ is always non-negative, it follows the upper bound of $\Sigma_\tau$:
\begin{equation}
    \Sigma_\tau \le D_{KL}(p_0^B || p_\infty^B), \label{eq:upperbound}
\end{equation}
and the equality is attained in the limit $\tau \to \infty$.

For the lower bound, we consider the difference
\begin{equation}
\Delta = \Sigma_\tau - D_{KL}(p_0^B || p_\tau^B)
= \sum_{j\in\{0,1\}}
(p_{0,j}^B - p_{\tau,j}^B)
\ln\frac{p_{\tau,j}^B}{p_{\infty,j}^B},
\label{eq:Delta}
\end{equation}
where $p_{t,j}^B$ denotes the probability of the bit being in state $j$ at interaction time $t$. Since $p_0'$ lies between $p_0$ and $p_\infty$, the factors
$(p_{0,j}^B - p_{\tau,j}^B)$ and
$\ln({p_{\tau,j}^B}/{p_{\infty,j}^B})$
always have the same sign, implying $\Delta \ge 0$.
The equality holds when $p_0=p_0'$ or $p_0'=p_\infty$, corresponding respectively to the cases of no interaction and $\tau\to\infty$.
We thus obtain the lower bound
\begin{equation}
D_{KL}(p_0^B || p_\tau^B) \le \Sigma_\tau,
\label{eq:lowerbound}
\end{equation}
with equality attained at $\tau=0$ or in the limit $\tau\to\infty$.

Combining Eq.~\eqref{eq:upperbound} and Eq.~\eqref{eq:lowerbound}, we recover the [Eq.(6)] in the main text. These inequalities suggest that the entropy production is bounded by the initial information-theoretic distance to the steady state.

\section{Asymptotic Behavior of $\Sigma_\tau$ near the Steady State}

In this section, we rigorously examine the relaxation dynamics and demonstrate that  the dissipation $\Sigma_\tau$ approaches $\Sigma_\infty$ exponentially in time.

To analyze this relaxation behavior, we start from the propagator $e^{\mathbf{R}\tau}$ of the composite system during the interaction interval $\tau$, where $\mathbf{R}$ is given by:
\begin{align*}
\mathbf R &= 
\left(
\begin{array}{cccc}
     -\gamma  (\sigma +1) & \gamma  (1-\sigma ) & 0 & 0 \\
     \gamma  (\sigma +1) & -\gamma  (1-\sigma )+\omega -1 & \omega +1 & 0 \\
     0 & 1-\omega  & -\gamma  (\sigma +1)-\omega -1 & \gamma  (1-\sigma ) \\
     0 & 0 & \gamma  (\sigma +1) & -\gamma  (1-\sigma ) \\
    \end{array} 
    \right).
    \label{eq:Rmatrix}
\end{align*} 
Then we apply spectral decomposition to $\mathbf{R}$ and the propagator can be expressed  as 
\begin{equation}
e^{\mathbf{R}\tau} = \sum_{i=0}^{3} e^{\lambda_i \tau} \frac{\mathbf{v}_i \mathbf{w}_i^\top}{ \mathbf{w}_i^\top \mathbf{v}_i},
\label{eq:spectral decomposition}
\end{equation}
in terms of  eigenvalues $\{\lambda_i\}$ of $\mathbf{R}$ and  corresponding sets of right and left eigenvectors $\{\mathbf{v}_i, \mathbf{w}_i\}$, where
\begin{equation}
    \lambda_0 = 0,\quad \lambda_1 = -2 \gamma,\quad
    \lambda_2 = -\alpha-\gamma -1 ,\quad
    \lambda_3 =  \alpha-\gamma -1,
    \label{eq:lambdas}
\end{equation}
with $\alpha = \sqrt{\gamma^2 + 2\gamma\sigma\omega + 1}$. The corresponding right eigenvectors, given by $\mathbf{R} \mathbf {v_i} = \lambda_i \mathbf{v_i}$, are
\begin{equation}
\begin{aligned}
\mathbf{v_0} &=
     (-\frac{(\sigma -1)^2 (\omega +1)}{(\sigma +1)^2 (\omega -1)},~
     -\frac{-\sigma \omega-\sigma+\omega+1}{(\sigma +1)(\omega -1)},~
     -\frac{\sigma -1}{\sigma +1},~1)^\top, \\
\mathbf{v_1} &=
     (-\frac{\omega +1}{\omega -1},~
     -\frac{-\omega -1}{\omega -1},~-1,~1)^\top, \\
\mathbf{v_2} &=
     (-\frac{1-\sigma}{\sigma +1},~
     -\frac{-\alpha+\gamma  \sigma -1}{\gamma  (\sigma +1)},~
     -\frac{\alpha+\gamma  \sigma +1}{\gamma  (\sigma +1)},~1)^\top ,\\
\mathbf{v_3} &=
     (-\frac{1-\sigma}{\sigma +1},~
     -\frac{\alpha+\gamma  \sigma -1}{\gamma  (\sigma +1)},~
     -\frac{-\alpha+\gamma  \sigma +1}{\gamma  (\sigma +1)},~1)^\top,
\end{aligned}
\end{equation}

and the corresponding left eigenvectors, given by $\mathbf{w_i}^\top \mathbf{R} = \lambda_i \mathbf{w_i}$, are
\begin{equation}
\begin{aligned}
\mathbf{w_0} &=
    (1,~1,~1,~1)^\top, \\
\mathbf{w_1} & =
    \left(
    \frac{(\sigma +1)^2}{(\sigma -1)^2},~
    -\frac{-\sigma -1}{\sigma -1},~
    -\frac{-\sigma -1}{\sigma -1},~1
    \right)^\top, \\
\mathbf{w_2} & =
    \left(
      -\frac{\sigma  \omega -\sigma +\omega -1}{(\sigma -1) (\omega +1)},~
      -\frac{-\omega  \alpha+\alpha+\gamma  \sigma  \omega -\gamma  \sigma -\omega +1}{\gamma  (\sigma -1) (\omega +1)},
      ~-\frac{-\alpha-\gamma  \sigma -1
      }{\gamma  (\sigma -1)},~1
      \right)^\top, \\
\mathbf{w_3} & =
    \left(-\frac{\sigma  \omega -\sigma +\omega -1}{(\sigma -1) (\omega +1)},~
      -\frac{\omega  \alpha -\alpha+\gamma  \sigma  \omega -\gamma  \sigma -\omega +1}{\gamma  (\sigma -1) (\omega +1)},
      ~-\frac{\alpha-\gamma  \sigma -1
      }{\gamma  (\sigma -1)},~1\right)^\top.
\end{aligned}
\end{equation}

With the definitions and tools established above, we now proceed to analyze the dynamics of the joint probability distribution. The joint probability distribution at the end of an interaction interval satisfies
\begin{equation}
    \mathbf{p}^{BD}_\tau = e^{\mathbf{R}\tau}\mathbf{p}_0^{BD}.
\end{equation}
Applying the spectral decomposition Eq.~\eqref{eq:spectral decomposition}, the joint distribution can be written as
\begin{align}
\mathbf{p}^{BD}_\tau
 = \sum_{i=0}^{3} e^{\lambda_i \tau} \frac{\mathbf{v}_i \mathbf{w}_i^\top}{\mathbf{w}_i^\top \mathbf{v}_i} \mathbf{p}_0^{BD} = \sum_{i=0}^{3} \mathbf{C}_i e^{\lambda_i \tau},
\end{align}
where $\mathbf{C}_i =  \mathbf{v}_i \mathbf{w}_i^\top \mathbf{p}_0^{BD}/(\mathbf{w}_i^\top \mathbf{v}_i).$ The outgoing bit distribution is obtained from the joint distribution through the projection operator $\mathcal{P}^B$,
\begin{align}
\mathbf{p}^B_\tau = \mathcal{P}^B \mathbf{p}^{BD}_\tau  = \sum_{i=0}^{3} \mathcal{P}^B \mathbf{C}_i e^{\lambda_i \tau} = \sum_{i=0}^{3} \mathbf{K}_i e^{\lambda_i \tau},
\end{align}
where the coefficients and projection operator are 
\begin{equation}
\begin{split}
\mathbf{K}_i = \mathcal{P}^B \mathbf{C}_i,  \quad
\mathcal{P}^B =
\begin{pmatrix}
1 & 1 & 0 & 0 \\
0 & 0 & 1 & 1
\end{pmatrix}.
\end{split}
\end{equation}

Since $\alpha<\gamma+1$, one obtains $\lambda_{1,2,3}<0$. Therefore, $\lambda_0 = 0$ corresponds to the stationary mode and the relaxation dynamics is determined by the nonzero eigenvalues. To identify the dominant relaxation mode, we compare the other eigenvalues:
\begin{equation}
\begin{gathered}
\lambda_3 - \lambda_1 = \alpha - (\gamma - 1) = \sqrt{\gamma^2 + 2\gamma\sigma\omega + 1} - \sqrt{\gamma^2 - 2\gamma + 1}  >0,  \\
\lambda_3-\lambda_2=2\alpha>0.
\end{gathered}
\end{equation}
Therefore, $\lambda_3$ has the largest real part among the nonzero eigenvalues and thus governs the long-time relaxation dynamics. In the long-time limit ($\tau \to \infty$), the contributions associated with faster-decaying modes corresponding to $\lambda_1$ and $\lambda_2$ become negligible, yielding
\begin{equation}
\mathbf{p}^B_\tau
=
\mathbf{p}_\infty^B
+
\delta \mathbf{p}^B_\tau,
\end{equation}
where
\begin{equation}
\delta \mathbf{p}^B_\tau \propto e^{-|\lambda_3| \tau}, \quad \sum_j \delta p^B_{\tau,j} = 0.
\end{equation}

With the dynamics fully characterized, we now analyze the behavior of $\Sigma_\tau$ in the long-time limit and define the dissipation gap $\Delta \Sigma$ as the difference between the entropy production and its upper bound:
\begin{equation}
    \Delta \Sigma =D_{KL}(p_0^B || p_\infty^B) - \Sigma_\tau=     D_{KL}(p_\tau^B || p_\infty^B).
    \label{eq:Delta_Sigma}
\end{equation}
To determine the asymptotic scaling of this gap, we perform a Taylor expansion of the KL divergence around $p^B_{\infty}$ for small deviations $\delta p^B_\tau$, 
\begin{equation}
\begin{split}
D_{KL}(p_\tau^B || {p}_\infty^B)  
&= \sum_{j\in\{0, 1\}} ({p}_{\infty,j}^B + \delta p_{\tau,j}^B) \ln\left(1 + \frac{\delta p_{\tau,j}^B}{{p}_{\infty,j}^B}\right)= \sum_{j\in\{0, 1\}} \delta p_{\tau,j}^B + \sum_j \frac{(\delta p_{\tau,j}^B)^2}{2{p}_{\infty,j}^B} + O[(\delta p_{\tau}^B)^3] \\
&=\sum_{j\in\{0, 1\}} \frac{(\delta p_{\tau,j}^B)^2}{2{p}_{\infty,j}^B} + \mathcal O[(\delta p_{\tau}^B)^3] \propto e^{-2|\lambda_3| \tau}.
\label{eq:asymptotic_2lambda}
\end{split}
\end{equation}
This result reveals that the dissipation gap $\Delta \Sigma$ vanishes exponentially at twice the rate of the bit distribution's relaxation.

\section{Solving for $\Theta(\tau)$}
% concrete form cited from ~\cite{2013PRLMQJ}

In this section, we analyze the function $\Theta(\tau)$, which characterizes the degree of relaxation towards thermal equilibrium. Following the exact solution structure of the MQJ model~\cite{2013PRLMQJ}(denoted as $\eta$ therein), $\Theta(\tau)$ can be expressed in the compact form %We start from the exact expression of $\Theta(\tau)$, formulated as the ratio of time-dependent auxiliary functions~\cite{2013PRLMQJ}
\begin{equation}\Theta(\tau) = \frac{\nu_2(\tau)P(\tau) + \nu_3(\tau)Q(\tau)}{P(\tau) + Q(\tau)}.\label{eq:Theta_PQ}
\end{equation}
The structural components $P(\tau)$ and $Q(\tau)$ are given by
\begin{align}P(\tau) &= \mu_2\Big[\mu_4\nu_3(\tau) + \mu_1\nu_1(\tau)\Big],\qquad Q(\tau) = \mu_3\Big[\mu_4\nu_2(\tau) + \mu_1\nu_1(\tau)\Big].
\end{align}
Here, the temporal evolution is governed by the exponential decay functions $\nu_i(\tau) = 1 - e^{-\lambda_i\tau}$ ($i=1,2,3$), with the respective relaxation rates $\lambda_i$ defined in Eq.~\eqref{eq:lambdas}. The constant parameters $\mu_j$ dictating the state-space weights are defined as
\begin{equation}
\begin{split}
\mu_1 = (\delta+\sigma)\omega,\quad\mu_2 = \alpha+\gamma+\sigma\omega,\quad\mu_3 = \alpha-\gamma-\sigma\omega, \quad\mu_4 = 1-\delta\omega.
\end{split}
\end{equation}
To capture the initial relaxation dynamics, we consider the short-time limit ($\tau \to 0$) and perform a Taylor expansion up to the second order of $\tau$. First, the exponential functions are expanded directly in terms of their relaxation rates $\lambda_i$:
\begin{equation}\nu_i(\tau) = \lambda_i\tau - \frac{\lambda_i^2}{2}\tau^2 + \mathcal{O}(\tau^3).
\end{equation}
Substituting these into the expressions for $P(\tau)$ and $Q(\tau)$ yields their corresponding short-time series representations
\begin{equation}P(\tau) = p_1\tau + p_2\tau^2 + \mathcal{O}(\tau^3),\quad Q(\tau) = q_1\tau + q_2\tau^2 + \mathcal{O}(\tau^3),
\end{equation}
where the first- and second-order expansion coefficients are straightforwardly identified as
\begin{align} p_1 &= \mu_2\Big(\mu_4\lambda_3 + \mu_1\lambda_1\Big),\quad p_2 = -\frac{\mu_2}{2}\Big(\mu_4\lambda_3^2 + \mu_1\lambda_1^2\Big),\quad q_1 = \mu_3\Big(\mu_4\lambda_2 + \mu_1\lambda_1\Big), \quad q_2 = -\frac{\mu_3}{2}\Big(\mu_4\lambda_2^2 + \mu_1\lambda_1^2\Big).
\end{align}
Finally, by assembling these polynomial expansions back into Eq.~\eqref{eq:Theta_PQ}, we obtain the short-time asymptotic behavior of the relaxation function:
\begin{equation}
\Theta(\tau) = \kappa\tau + c_2\tau^2 + \mathcal{O}(\tau^3),
\label{eq:exp_of_Theta}
\end{equation}
where
\begin{equation}
\kappa = \frac{\lambda_2 p_1 + \lambda_3 q_1}{p_1 + q_1}, \quad
c_2 = \frac{2\lambda_2 p_2 - \lambda_2^2p_1 + 2\lambda_3 q_2 - \lambda_3^2q_1}{2(p_1 + q_1)}
- \frac{(\lambda_2 p_1 + \lambda_3 q_1)(p_2 + q_2)}{(p_1 + q_1)^2}.
\end{equation}

% where the effective initial relaxation rate $\kappa$ and the curvature coefficient $c_2$ are strictly determined by the algebraic combination of $\{p_j, q_j, \lambda_j\}$.

% To define $\kappa$ and $c_2$ explicitly, we introduce the denominator coefficients $d_1, d_2$ and the numerator coefficients $n_2, n_3$ from the expansion of Eq.~\eqref{eq:Theta_PQ}:
% \begin{align}
% d_1 &= p_1 + q_1, \quad d_2 = p_2 + q_2, \\
% n_2 &= \lambda_2 p_1 + \lambda_3 q_1, \quad n_3 = \lambda_2 p_2 - \frac{\lambda_2^2}{2}p_1 + \lambda_3 q_2 - \frac{\lambda_3^2}{2}q_1.
% \end{align}
% The effective initial relaxation rate $\kappa$ and the curvature coefficient $c_2$ are then strictly determined as:
% \begin{equation}
%  \kappa = \frac{n_2}{d_1} , \quad  c_2 = \frac{n_3}{d_1} - \frac{n_2 d_2}{d_1^2} .
% \label{eq:kappa_c2_final}
% \end{equation}

\section{The Crossover Time $\tau^*$ of the Eraser Regime}

In this section, we characterize a specific dynamical regime where the system transitions from a purely dissipative state to an information erasure state. We focus on the scenario where $\delta < 0 < \epsilon$. In this parameter space, the heat exchange $\Delta Q(\tau)$ remains strictly negative for all $\tau$~\cite{2013PRLMQJ}, while the bit entropy change $\Delta S_B(\tau)$ exhibits a sign reversal as $\tau$ increases. This phenomenon leads to the ``delayed eraser'' behavior, where the operational mode of the device is determined by the interaction duration.

\begin{figure}[htb]
    \centering
    \includegraphics[width=0.7\linewidth]{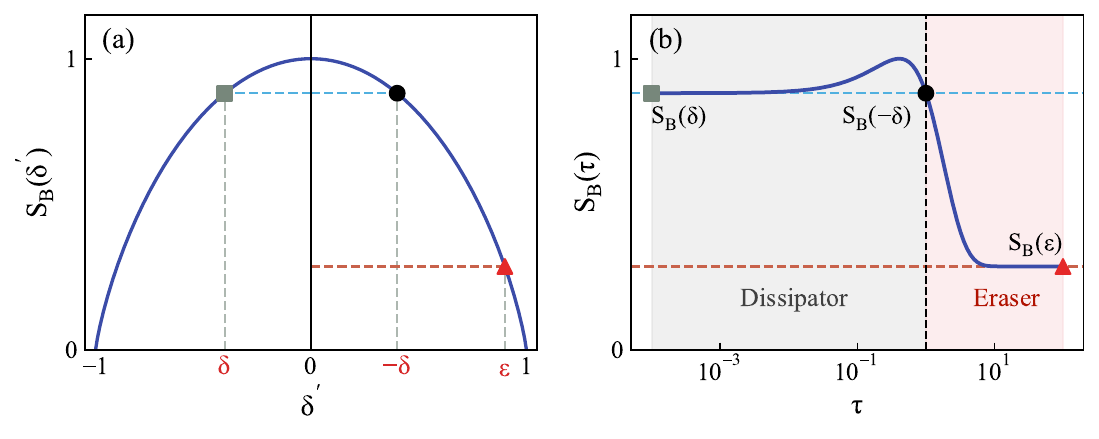}
    \caption{Entropy characteristics and evolution regimes.(a) Shannon entropy $S_B(\delta')$ as a function of the bias $\delta'$, illustrating even symmetry and strict concavity with a maximum at $\delta'=0$. The points $\delta$ (gray square), $-\delta$ (black circle), and $\epsilon$ (red triangle) are marked to show the states within the regime $|\epsilon| > |\delta|$. 
    (b) Evolution of $S_B$ as a function of interaction time $\tau$. The marked entropy levels corresponding to $\delta$, $-\delta$, and $\epsilon$ match the points in panel (a), illustrating the transition from the dissipator regime to the eraser regime. Gray and red shaded regions denote the dissipator and eraser regimes, respectively. Parameters are $\delta=-0.4$, $\epsilon=0.9$, $\omega=0.7$, and $\gamma=1$.}
    \label{fig:shannon entropy}
\end{figure}

The physical origin of this crossover lies in the symmetry of the entropy function and the monotonicity of the relaxation process. The Shannon entropy $S(\delta')$ is an even and concave function of $\delta'$, attaining its maximum at $\delta' = 0$ as shown in Fig.~\ref{fig:shannon entropy}(a). Consequently, the sign of the entropy production is entirely determined by the magnitude of the bias: $\operatorname{sign}\big(\Delta S_B\big)= \operatorname{sign}\!\big(|\delta| - |\delta'|\big).$  According to the supplemental material  of Ref.~\cite{2013PRLMQJ}, the system distribution evolves monotonically from the initial state to the steady state. Therefore, for a crossover to occur, the bias $\delta'(\tau)$ must evolve monotonically from its initial negative value $\delta < 0$ to a positive stationary value $\epsilon > 0$, with the additional requirement that $\epsilon > |\delta|$ as illustrated in Fig.~\ref{fig:shannon entropy}(b).

Based on these properties, we can now identify the precise condition at which the entropy change $\Delta S_B$ changes its sign. We define the crossover time $\tau^*$ as the critical moment at which the bit entropy change equals zero, $\Delta S_B(\tau^*) = 0$. This occurs when the outgoing bias reaches the symmetric counterpart of its initial value, 
\begin{equation}
\delta'(\tau^*) = -\delta.
\label{eq:tauStar_delta}
\end{equation}
Substituting the dynamical relation $\delta'(\tau) = \delta - (\delta - \epsilon)\Theta(\tau)$ into Eq.~\eqref{eq:tauStar_delta}, we obtain an exact condition for the relaxation function at the crossover point:
\begin{equation}
\Theta(\tau^*) = \frac{2\delta}{\delta-\epsilon} = \frac{2}{1-\epsilon/\delta}.
\label{eq:tauStar_Theta}
\end{equation}

This relation establishes a clear operational boundary for the system at fixed intrinsic parameters. 
For $\tau < \tau^*$, the system operates in the dissipative regime $(\Delta Q<0,\, \Delta S_B>0)$, 
whereas for $\tau > \tau^*$ it enters the information eraser regime $(\Delta Q<0,\, \Delta S_B<0)$.

\section{Condition for the Existence of a Maximum Erasure Power}

In this section, we investigate the optimization of the eraser power $P_E(\tau)$ with respect to the interaction duration $\tau$, and examine the possible occurrence of a finite-time maximum.

The possible emergence of a finite-time maximum is closely related to the distinct short- and long-time behaviors of the entropy change $\Delta S_B(\tau)$. Therefore, we begin with an analysis of the temporal evolution of $\Delta S_B(\tau)$. To capture the early-stage dynamics($\tau \to 0$),  we express the outgoing bit bias as $\delta'(\tau) = \delta + (\epsilon-\delta)\Theta(\tau)$ and perform a Taylor expansion of the Shannon entropy $S(\delta')$ around the initial bias $\delta$ :
\begin{equation}
\Delta S_B(\tau) = S'(\delta)\,(\epsilon-\delta)\Theta(\tau) + \frac{1}{2} S''(\delta)\,(\epsilon-\delta)^2\Theta(\tau)^2 + \mathcal{O}(\Theta^3).
\end{equation}
Using the derivatives of $S(\delta')$, $S'(\delta) = -\operatorname{artanh}(\delta)$ and $S''(\delta) = -1/(1-\delta^2)$, and inserting Eq.~\eqref{eq:exp_of_Theta}, we arrive at the short-time expansion for the entropy change. To facilitate the analysis of the eraser power $P_E(\tau)$ in later sections, we retain terms up to second order in $\tau$:
\begin{equation}
\begin{split}
\Delta S_B(\tau) &= -\operatorname{artanh}(\delta)\,(\epsilon-\delta)\left(\kappa\tau + c_2\tau^2\right)
 - \frac{(\epsilon-\delta)^2}{2(1-\delta^2)}\kappa^2\tau^2 
 + \mathcal{O}(\tau^3) \\
 &= -\operatorname{artanh}(\delta)\,(\epsilon-\delta)\kappa\tau - \left[\operatorname{artanh}(\delta)(\epsilon-\delta)c_2
+\frac{(\epsilon-\delta)^2}{2(1-\delta^2)}\kappa^2 \right] \tau^2 +\mathcal{O}(\tau^3).
\end{split}
\label{eq:dSB_smalltau}
\end{equation}

In the long-time limit($\tau \to \infty$), the system fully relaxes to the stationary state. Consequently, the outgoing bit bias reaches the stationary limit, $\delta_\infty = \epsilon$. The corresponding long-time limit for the entropy change is given by:
\begin{equation}
\Delta S_B(\infty) = S(\epsilon) - S(\delta).
\label{eq:dSB_infinitetau}
\end{equation}

With the asymptotic behaviors of  $\Delta S_B(\tau)$ established, we now turn to the behavior of the eraser power $P_E(\tau)$,which is given by [Eq.~(7)] in the main text, 
\begin{equation}
P_E(\tau) = -\frac{\Delta S_B(\tau)}{\tau}.
\end{equation}
From Eq.~\eqref{eq:dSB_smalltau}, $\Delta S_B(\tau)$ grows linearly with $\tau$ at short times, implying a finite nonzero initial value of the $P_E(\tau)$ as $\tau \to 0$. In contrast, $\Delta S_B(\tau)$ saturates to a constant at long times as given in Eq.~\eqref{eq:dSB_infinitetau}, leading to $P_E(\tau) \sim 1/\tau \to 0$ as $\tau \to \infty$. This transition from a finite short-time value to a vanishing long-time limit naturally suggests that $P_E(\tau)$ may attain a maximum at a finite interaction time $\tau$.

To determine whether $P_E(\tau)$ develops a maximum at finite $\tau$, we examine its initial slope via a short-time expansion,
\begin{equation}
P_E(\tau) = A_0 + A_1 \tau + \mathcal{O}(\tau^2),
\label{eq:PESmallTau}
\end{equation}
where the coefficients are obtained from the short-time expansion of $\Delta S_B(\tau)$ in Eq.~\eqref{eq:dSB_smalltau},
\begin{equation}
    A_0 = \operatorname{artanh}(\delta)(\epsilon-\delta)\kappa, \quad
    A_1 =
\operatorname{artanh}(\delta)(\epsilon-\delta)c_2
+\frac{(\epsilon-\delta)^2}{2(1-\delta^2)}\kappa^2.
\end{equation}
In particular, $A_0  > 0$ gives the finite initial value of the power, while the linear coefficient $A_1$ governs the subsequent evolution. If $A_1 > 0$, the power initially increases with $\tau$ before eventually decaying to zero at long times, thereby guaranteeing the existence of a maximum at finite $\tau$. This provides a sufficient criterion for the emergence of an optimal eraser power, which can be expressed in the dimensionless form
\begin{equation}
\frac{c_2}{\kappa^2} > -\frac{\epsilon-\delta}{2(1-\delta^2)\operatorname{artanh}(\delta)}.
\label{eq:bcriterion}
\end{equation}

\section{Refrigerator Regime and Performance Limits}
\label{Refrigerator}

In this section, we analyze the refrigeration regime and its performance. 
The regime is defined by heat extraction from the cold reservoir ($\Delta Q > 0$), accompanied by an increase in the informational entropy of the bit stream ($\Delta S_B > 0$).

From [Eq.~(7)] in the main text, the cooling power and efficiency are given by
\begin{equation}
P_R(\tau) = \frac{\Delta Q(\tau)}{\tau}, \quad \eta_R(\tau) = \frac{\Delta\beta\Delta Q(\tau)}{\Delta S_B(\tau)}. 
\end{equation}
To proceed, we examine the dynamical behavior of the heat exchange $\Delta Q(\tau)$ and $\Delta S_B(\tau)$, which fully determines $P_R(\tau)$ and $\eta_R(\tau)$. Having established the behavior of $\Delta S_B(\tau)$ in Eqs.~\eqref{eq:dSB_infinitetau} and \eqref{eq:dSB_smalltau}, we now focus on dynamics of the heat exchange $\Delta Q(\tau)$. We consider the short-time limit ($\tau \to 0$). Substituting Eq.~\eqref{eq:exp_of_Theta} into $\Delta Q(\tau) = \Delta E (\delta - \epsilon)\Theta(\tau)/2$, we have
\begin{equation}
\begin{split}
\Delta Q(\tau) 
&= \frac{\Delta E}{2}(\delta-\epsilon)\left[\kappa\tau + c_2\tau^2 + \mathcal{O}(\tau^3)\right] \\
&= \frac{\Delta E}{2}(\delta-\epsilon)\kappa\tau + \mathcal{O}(\tau ^2).
\end{split}
\label{eq:dQ_smalltau}
\end{equation}
In the long-time limit($\tau \to \infty$),  since the system reaches its stationary state, the heat exchange is 
\begin{equation}
\Delta Q(\infty) = \frac{\Delta E}{2}(\delta-\epsilon).
\label{eq:dQ_infinitetau}
\end{equation}

\begin{figure}[htb]
    \centering
    \includegraphics[width=1\linewidth]{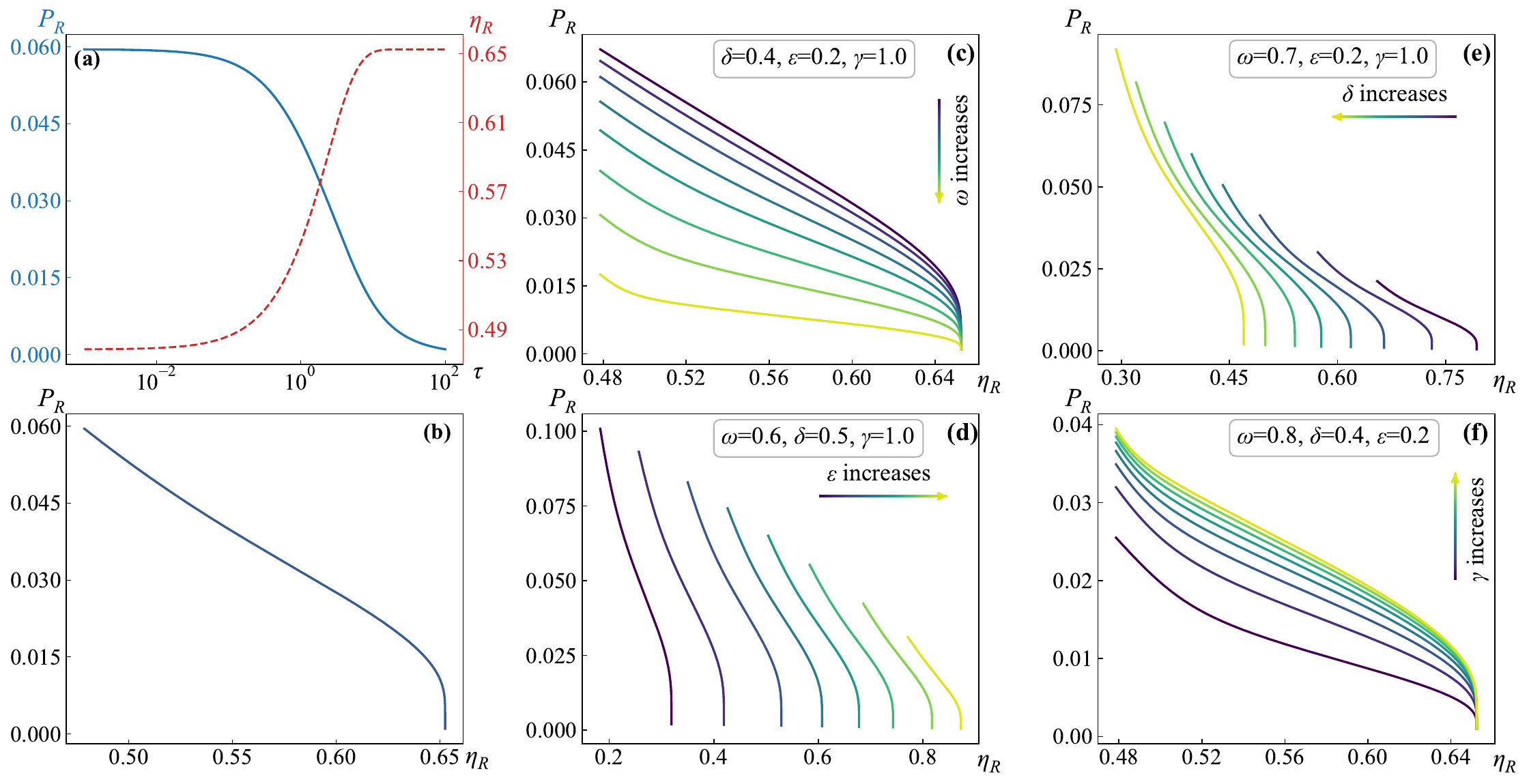}
    \caption{
    (a) Refrigeration power $P_R$ (solid line) and efficiency $\eta_R$ (dashed line) as functions of the interaction time $\tau$. (b) Trade-off between $P_R$ and $\eta_R$. The power decreases monotonically with increasing $\tau$ and vanishes in the long-time limit, while the efficiency increases monotonically and approaches its asymptotic value. Parameters are $\delta = 0.4$, $\epsilon = 0.2$, $\omega = 0.5$, and $\gamma = 1.0$.
    (c-f) Parameter dependence of the refrigeration performance in the trade-off between $P_R$ and $\eta_R$. Panels (c)-(f) show systematic sweeps of (c)$\omega =\{0.3,0.39,0.47,0.56,0.64,0.73,0.81,0.9\} $, $\epsilon =\{0.1, 0.14, 0.19, 0.23, 0.27, 0.31, 0.36, 0.4\}$, $\delta =\{0.3,  0.34, 0.39, 0.43, 0.47, 0.51, 0.56, 0.6\}$, and $\gamma =\{0.5, 1, 1.5, 2, 2.5, 3, 3.5, 4\}$, respectively, while the remaining parameters are fixed at the values indicated in each panel.
    }
    \label{fig:tradeoff of friger}
\end{figure}

With the detailed analysis of $\Delta S_B(\tau)$ and $\Delta Q(\tau)$, we now return to the analysis of the refrigeration performance quantities $\eta_R(\tau)$ and $P_R(\tau)$. In the short-time limit, using the perturbative expansions of $\Delta Q(\tau)$ and $\Delta S_B(\tau)$ given in Eqs.~\eqref{eq:dQ_smalltau} and \eqref{eq:dSB_smalltau}, we obtain
\begin{equation}
P_R(\tau \to 0) = \frac{\Delta E}{2}(\delta-\epsilon)\kappa, \quad 
\eta_R(\tau \to 0) = \frac{\operatorname{artanh}(\epsilon)}{\operatorname{artanh}(\delta)}.
\label{eq:P_eta_smalltau}
\end{equation}
In the long-time limit ($\tau \to \infty$), the total extracted heat and the entropy change of the bit saturate according to Eqs.~\eqref{eq:dSB_infinitetau} and ~\eqref{eq:dQ_infinitetau}. Consequently, the time-averaged refrigeration power vanishes, while the corresponding efficiency remains finite:
\begin{equation}
P_R(\infty) = 0, \quad 
\eta_R(\infty) = \frac{\operatorname{artanh}(\epsilon)(\delta-\epsilon)}{S(\epsilon)-S(\delta)}.
\label{eq:P_eta_infinitetau}
\end{equation}
The above asymptotic analysis reveals a clear trade-off between power and efficiency in the two limiting regimes: while the efficiency remains finite in the long-time limit, the power vanishes; conversely, the short-time regime yields finite power but generally lower efficiency.

We further investigate the optimization of the refrigeration performance. Within the parameter range scanned in this work, no regime is found where the refrigeration power and efficiency increase simultaneously. Instead, both quantities exhibit monotonic behavior, as shown in Fig.~\ref{fig:tradeoff of friger}. In particular, the refrigeration power decreases from its initial value in Eq.~\eqref{eq:P_eta_smalltau} as the interaction time $\tau$ increases, gradually approaching zero in the long-time limit, while the efficiency shows the opposite trend and saturates to its asymptotic value in Eq.~\eqref{eq:P_eta_infinitetau}, as illustrated in Fig.~\ref{fig:tradeoff of friger}(a). The corresponding $P_R$–$\eta_R$ trade-off curve in Fig.~\ref{fig:tradeoff of friger}(b) further confirms this monotonic correspondence. When system parameters are varied, as shown in Fig.~\ref{fig:tradeoff of friger}(c)–(f), the trade-off curves are deformed but preserve their overall monotonic character. Specifically, variations in $\omega$, $\epsilon$, $\delta$, and $\gamma$ primarily lead to shifts and stretching of the $P_R$–$\eta_R$ profiles, effectively tuning the accessible operating region, but no non-monotonic features are observed. 

\bibliography{refs}